# Beyond Cortisol! Physiological Indicators of Welfare for Dogs: Deficits, Misunderstandings and Opportunities

Cobb, M.L.[1*], Jiménez, A.G.[2], Dreschel, N.A.[3]

[1] The University of Melbourne, Animal Welfare Science Centre, Melbourne Veterinary School, Faculty of Science, VIC Australia https://orcid.org/0000-0001-5735-5126
*Corresponding author: mia.cobb@unimelb.edu.au

[2] Colgate University, Department of Biology, Hamilton, NY United States
https://orcid.org/0000-0001-9586-2866

[3] Pennsylvania State University, Department of Animal Science, University Park, PA United States
https://orcid.org/0000-0003-4258-8255

**Abstract**

This paper aims to initiate new conversations about the use of physiological indicators when assessing the welfare of dogs. There are significant concerns about construct validity - whether the measures used accurately reflect welfare's complexity. The goal is to provide recommendations for future inquiry and encourage debate. We acknowledge that the scientific understanding of animal welfare has evolved and bring attention to the shortcomings of commonly used biomarkers like cortisol. These indicators are frequently used in isolation and with limited salient dog descriptors, so fail to reflect the canine experience adequately. Using a systems approach, we explore various physiological systems and alternative indicators, such as heart rate variability and oxidative stress, to address this limitation. It is essential to consider factors like age, body weight, breed, and sex when interpreting these biomarkers correctly, and researchers should report on these in their studies. This discussion identifies possible indicators for both positive and negative experiences. In conclusion, we advocate for a practical, evidence-based approach to assessing indicators of canine welfare, including non-invasive collection methods. We acknowledge the complexity of evaluating experiential responses in dogs across different situations and the need for continued work to improve practices and refine terminology. This will enhance our ability to accurately understand welfare and improve the wellbeing of dogs, serving to inform standards of animal welfare assessment. We hope this will promote more fundamental research in canine physiology to improve construct validity, leading to better practices, ultimately improving the lives of dogs.

**Introduction**

Animal wellbeing has become increasingly important as scientific understanding and modern community attitudes have evolved. Recent evidence confirms the sentience of many species living with people globally, including domestic dogs (*Canis familiaris*) (Browning and Birch, 2022; Browning and Veit, 2023). We find dogs in diverse roles, as human companions, protectors, guides, herders, detectors, free ranging alongside us, and as a focus for entertainment and gambling. This understanding of sentience emphasises that dogs value a 'good life' of agency and positive mental experiences (Littlewood et al., 2023). With this comes a moral obligation for those involved in the regulation, care and management of dogs (Mackenzie, et al., 2014; Cobb et al., 2021). It is no longer enough to regulate that dogs' experiences are free from neglect, suffering, cruelty or harm; people should ensure that our canine companions and co-workers can lead a good life. Specifically, we should reduce or remove avoidable negative experiences (e.g., hunger, pain, exposure to extreme heat/cold, isolation, etc.) and strive to provide dogs with positive experiences (e.g., satiety, prompt veterinary care, comfort, social connection, etc.) (Webster, 2016; Lawrence et al., 2019; Mellor et al., 2020). Assurance of wellbeing will underpin the social license to operate in many interactions with dogs in working and companion roles (Hampton et al., 2020). Industries and practices that depend upon dogs will need to assure the wider community of this to be sustainable in the future (Cobb et al., 2021). While behavioural and physiological indicators give us important information about how dogs experience their lives (Polgár et al., 2019), current practices in canine welfare science don't align with the evidence base. Some of the regularly used physiological measures are being used and reported in ways that are problematic. This discussion paper examines the use of physiological indicators in dog welfare assessment, evaluating current limitations, misconceptions, gaps and opportunities for future research.

Animal welfare assessment and assurance to safeguard dogs rely on many factors, including resources (including monetary budget, personnel, skills and equipment), transparency, traceability, and validated indicators of negative experiences (commonly referred to as stress, distress or suffering) and positive experiences (understood as wellbeing). These include observations of behaviour and physiology. However, monetary budgetary constraints, evidence deficits, and misunderstandings within canine science have resulted in assessment practices that do not align with the evidence base or lack the evidence upon which to base good practice. In particular, there are significant concerns about construct validity - whether these physiological indicators actually measure what we claim they measure about dog welfare. There is a clear need (as identified relevant to animal welfare science more broadly by Mason, 2023) to explore the canine physiological indicators relating to affective states (negative and positive, see Mendl et al., 2022) that are used to assess welfare. Unlike a traditional review, this perspective paper intends to stimulate debate and drive progress within the field. This perspective builds on the existing canine science evidence base and cross-species physiological research to identify knowledge gaps and novel approaches in dog welfare assessment. We aim to advance the use of validated physiological indicators to meet the emerging need to centre dog wellbeing in research, regulation and human-animal interactions.

**Aligning the meaning of 'animal welfare' across multiple disciplines**

Modern animal welfare science understands 'animal welfare' as representing the full range of mental experiences in an animal's life, which is complex and cannot be directly measured (Mason & Mendl, 1993). This view recognises that animals, including dogs, have essential affective (emotional), social, and cognitive needs beyond basic requirements (De Winkel et al., 2024; Flint et al., 2024). Providing animals with agency and opportunities for play, companionship, safety, and a diverse range of behavioural interactions creates positive experiences. These contribute to 'a good life', one where the balance of experiences, when considered over time, is positive. Individual variation across dimensions like personality, cognitive bias, prior experience, and social attachment can all influence how dogs may have different mental experiences (i.e. different welfare) in response to the same social or environmental conditions.

The Five Domains Model of Animal Welfare (Mellor et al., 2020) provides a framework widely used in animal welfare science and its translation to care provision and regulation (e.g., Beausoleil, et al., 2023). It assessed indicators of good and bad effects that animals experience across the four physical and functional domains: 1. Nutrition, 2. Physical Environment, 3. Health and 4. Behavioural Interactions (with other animals, people and the environment). When considered together at any point in time, these inform an animal's Mental Experience (domain 5), which represents its welfare. Positive experiences such as feeling comfortably satiated after eating, enjoying the sensation of lying in a soft bed, feeling strong while running fast and playing happily with their family all contribute to positive (sometimes referred to as good) animal welfare and wellbeing. However, feeling thirsty, uncomfortable from lying on wet concrete, having itchy skin from an untreated allergy and feeling lonely from social isolation are unpleasant. Such negative experiences contribute to poor animal welfare and suffering. In reality, most dogs live with a mix of these experiences, and so investigating their welfare requires us to consider additional aspects of their experiences such as duration, intensity and significance to the individual. When considered over time, a life that has a balance of neutral or positive experiences (accepting that some negative experiences from time to time are inevitable) is considered a life worth living, and one that balances as positive is a good life (Lawrence et al., 2019).

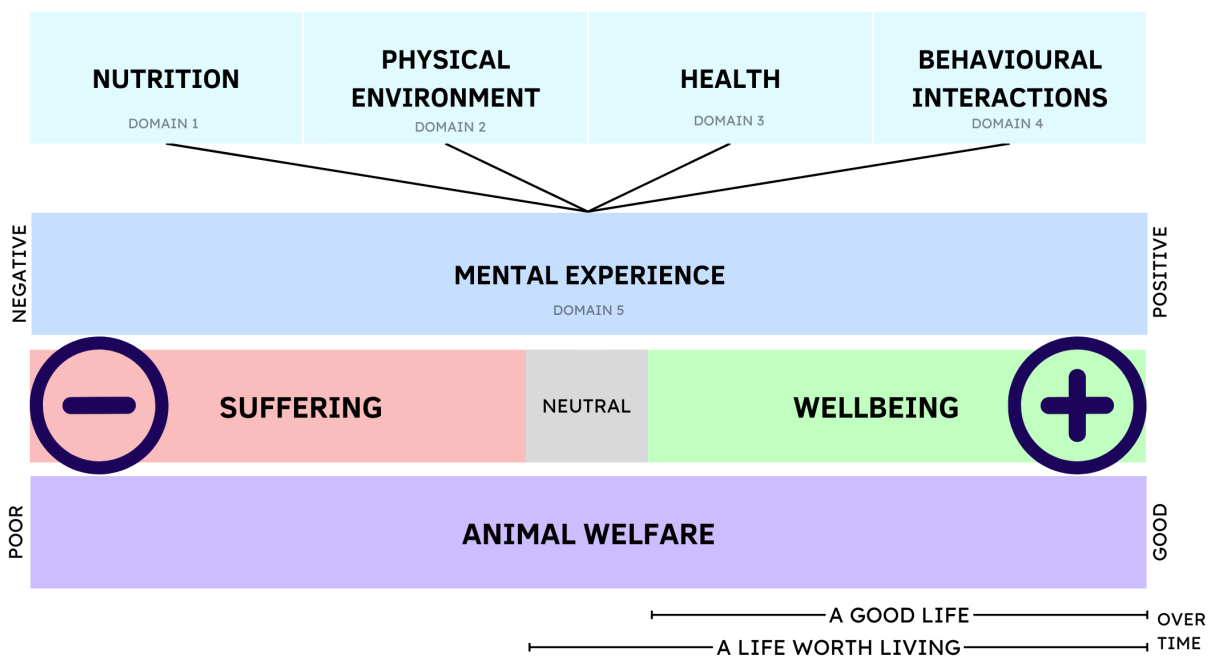

Figure 1. Conceptual alignment of an animal's welfare with constructs of suffering, wellbeing, a life worth living, a good life, animal mental experience and the Five Domains of Animal Welfare model.

**Adopting precise stress terminology in dog welfare research**

Stress is a key biological concept, with the word widely applied across physiological, psychological, social, medical and environmental fields over the last fifty years (Lu et al., 2021). During this period, our understanding of the stress concept has evolved significantly and is expected to continue developing. Referring to *stress* in animal welfare science can be confusing, as the word might be used to reference many relevant terms (stimulus, effect, affective state, etc.). Here, we follow the terminology of Lu and colleagues (2021), defining *stress* as "a state of homeostasis being challenged", and *stressors* as "factors with the potential to directly challenge homeostasis" (Figure 2). Homeostatic challenges vary in intensity and duration; a strong challenge may result from an intense and acute stressor, or when repeated (singular or multiple) low-to-moderate stressors are experienced persistently.

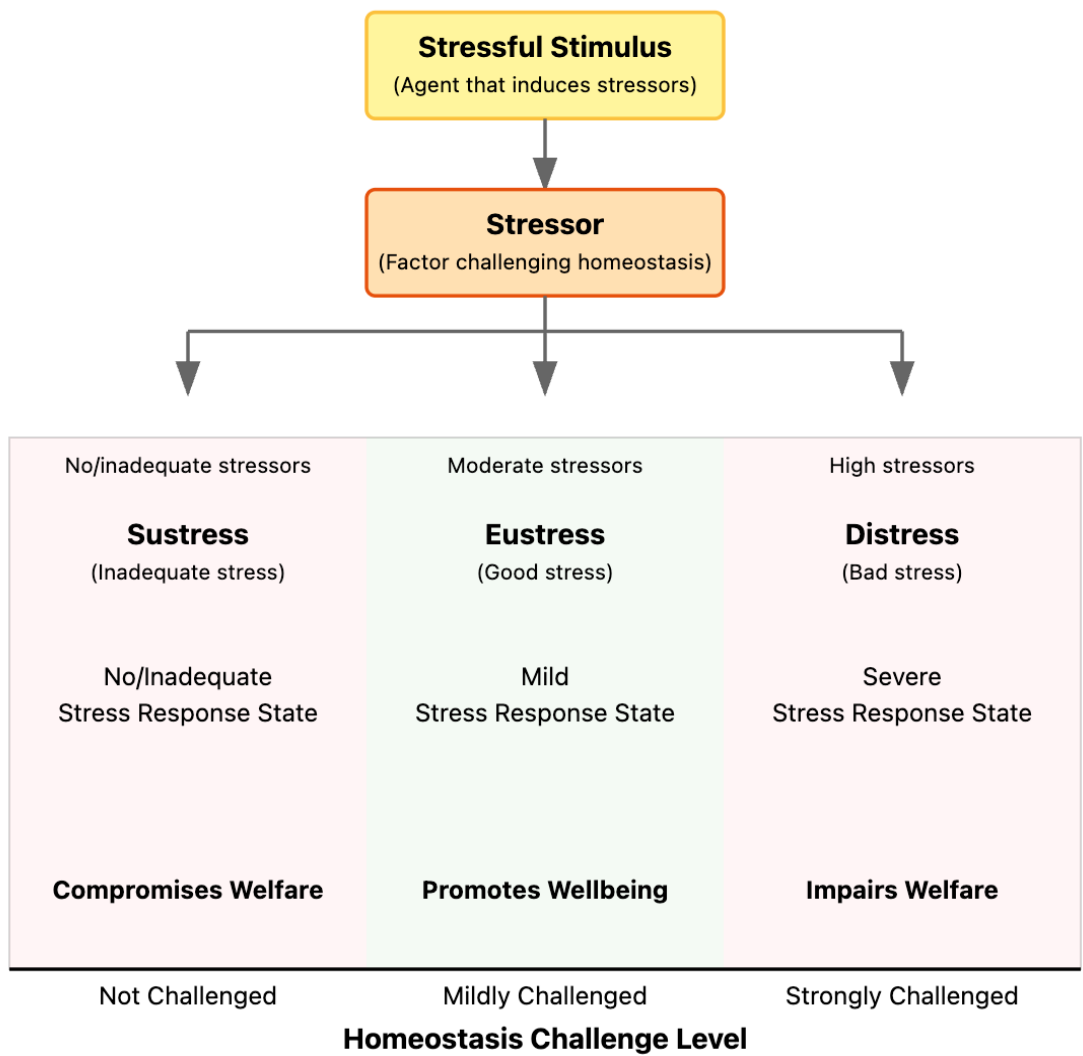

*Figure 2. Conceptualisation of key physiological stress-related terms in the stress system response, following terminology from Lu et al., 2021.*

Using 'stress' as a blanket term oversimplifies dogs' emotional and mental states, no longer aligning with the current scientific understanding of animal welfare. Although widely used, this term can potentially lead to misinterpretations when evaluating canine welfare. Instead, precise terminology should be used to improve the accuracy of our assessments, reflecting the specific negative and positive affective experiences of dogs. Mellor and colleagues (2020) categorise aversive experiences as either survival-critical negative affects, or situation-related negative affects. Survival-critical negative affects are the animal's sensory response to "imbalances or disruptions in the *internal* physical/functional state of animals" (Mellor et al., 2020). These are linked to inherent biological mechanisms that stimulate or relate to behaviours critical to the animal's survival, specific to the resolution of that experience. Survival-critical affects include breathlessness triggering increased respiration, hunger prompting food-seeking and eating, pain causing avoidance behaviours, and thirst driving water-seeking and drinking. These behaviours subside once the functional/physical problem has been met. When animals cannot effectively respond to aversive experiences (e.g., being tethered in the hot sun for a week), welfare impact is greater than from acute but brief experiences (e.g., hearing thunder twice in five minutes). In contrast, situation-related negative affects are derived from the animal's perception of its *external* circumstances. Examples include frustration, loneliness, fear, hypervigilance and depression. These categories may overlap - a dog with a broken leg experiences both survival-critical affects (pain leading to avoiding movement of the injured limb, whimpering and guarding the injured area) and situation-related affects (fear of unfamiliar hospital environment and loneliness while away from social attachment figures). Addressing the pain, fear, and social isolation (both types of affects) is important for welfare improvement.

Objective animal-based evidence is needed to infer welfare-relevant affects. For example, a dog's awake inactivity could indicate comfortable relaxation or fear-induced withdrawal and immobility. Accurate assessment requires the consideration of multiple factors: species-typical behaviours (e.g., ear position, mouth tension, tail movement, sleep-activity budget), environmental conditions (e.g., temperature, loud noises, familiarity with surroundings) and physiological indicators (e.g., breathing patterns, heart rate variability). These combined sources of evidence help identify both negative and positive affective states, informing a welfare assessment (Fureix & Meagher, 2015). Identifying different stress states (*sustress*, *eustress* [considered 'stimulation', per Broom, 2017], or *distress*; Figure 2), requires recognising the multiple coordinated physiological systems in the individual animal (Figure 3). These systems maintain homeostasis through complex feedback loops, with disruption in one system triggering responses within that localised system and among the other systems, to bring the organism back to a stable state. The endocrine, nervous, and immune systems work in conjunction with the gastrointestinal, integumentary, cardiovascular and circulatory systems to restore and maintain stability. This interconnected quality highlights the risks of relying on single measures, like cortisol alone, provides insufficient evidence of stress or welfare. It appears there are still fundamental deficits in our understanding of canine physiology in welfare science, and opportunities presented by new methodologies.

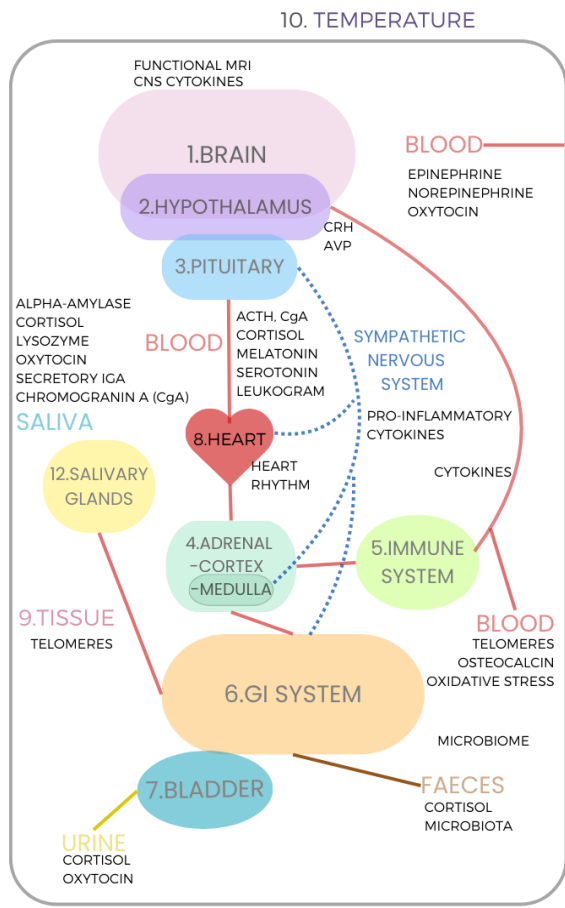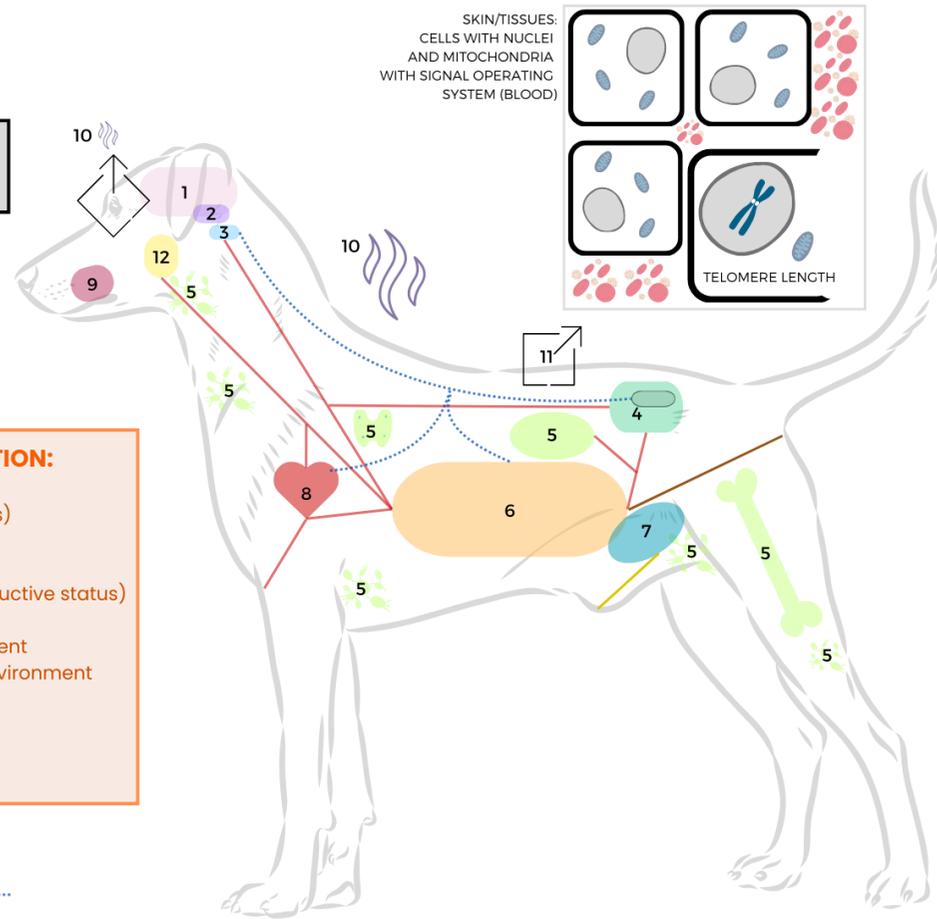

Figure 3. The systems and indicators of experiential physiological responses in dogs.

**A systems approach to stress physiology and behaviour**

The nervous system determines whether a stress response is initiated based on individualised perception of stressors (Figure 3, 'Individual variation'). This depends on many factors, including sensory capabilities, genetics, health, and prior experiences. A dog's sensory perception of the world, including vision, hearing, olfaction, and touch, may determine what environmental conditions are perceived as stressors. Many of these features have genetic components (e.g., visual acuity, ear carriage, coat length). Even a dog's level of pain perception (sensitivity) may vary by breed (Caddiell et al., 2023) and personality (Lush and Ijichi, 2018). The presence of disease can also affect an animal's sensitivity to their environment. The limbic system processes these inputs, potentially activating the hypothalamus to mount a physiological stress response if homeostasis of the organism appears threatened. There is evidence that specific gene variants may play a role in this processing of a dog's fear response (Zapata et al., 2016). Physical environment (housing, temperature, humidity, nutrient availability, etc.) and social interactions (presence of conspecifics and other species, including humans) may also affect an individual's stress response (McEwen, 2001). Prior experiences may affect whether they perceive a specific stimulus as aversive. Previous exposure to stimuli influences response, as seen in dogs with thunderstorm phobia responding to changes in barometric pressure or distant thunder sounds, while other dogs remain unaffected (Dreschel and Granger, 2005).

When a stimulus registers as a stressor, it initiates a systemic stress response (physiological processes occurring systematically in the whole body, independent of where stressors originated from, as per Lu et al., 2021). This response operates through two primary pathways: the Sympathetic Nervous System (SNS) and the Hypothalamic-Pituitary-Adrenal (HPA) axis. As both systems have been well described (Chrousos, 1998), we include only a brief overview here to inform the discussion.

SNS activation begins when the hypothalamus signals through preganglionic neurons to the thoracolumbar grey column of the spinal cord, then to target organs via ganglionic axons, primarily using norepinephrine as a neurotransmitter (de Lahunta et al., 2021). This rapid, short-acting response is limited only by neural impulse speed and neurotransmitter (epinephrine and norepinephrine) actions at the widespread target organs. SNS responses could potentially be measured at many different locations and by its effects on other systems. The cardiovascular and circulatory systems respond with increased heart rate, pulse, and changes in blood pressure and heart rate variability. Increased respiratory rate and pupil dilation occur. The immune system (comprising thymus, spleen, lymph nodes and bone marrow, refer Figure 3) responds with changes in white blood cell distribution and immunoglobulin levels. The enteric system responds with decreased motility of the stomach and intestines, and gastrointestinal sphincter contraction (McCorry, 2007). Salivary alpha-amylase increases have been noted in humans (Nater et al., 2006). The skeletal system may respond to acute stress via osteocalcin release (Berger et al., 2019).

The HPA axis, highly conserved among vertebrate species (Bouyoucos et al., 2021), activates in response to acute stressors through a cascade of hormones: the hypothalamus causes release of corticotropin releasing hormone (CRH) which travels to the anterior pituitary gland. This causes release of adrenocorticotropic hormone (ACTH) from the anterior pituitary into the circulatory system, triggering the release of corticosteroids (i.e.

cortisol in canines) from the adrenal cortex into the circulating blood. The glucocorticoids have effects on multiple systems: immune, gastrointestinal, endocrine, and cardiovascular (McEwen, 2019). Communication occurs between the immune and endocrine systems and central nervous system via cytokines (Granger et al., 2003). There is evidence that the gut microbiota and permeability are also affected (Dinan and Cryan, 2012). This hormonal response (HPA activation) is relatively slower and longer acting than the SNS response even after the stressor has subsided and self-regulates through negative feedback, with glucocorticoids suppressing ACTH and CRH to restore homeostasis.

The duration of homeostatic disruption significantly influences physiological responses. Acute responses occur with specific events (e.g. new environments, frightening sounds, brief transportation) and help survival through short-term challenges. Continued exposure to a stimulus results in two possible outcomes: habituation (the animal learns to no longer react to the stimulus as a stressor), or chronic stress (develops from repeatedly or continuous exposure to a stressor). Notably, chronic stressors (e.g., inconsistent training, resource competition) do not need to be intense to present a sustained negative experience for dogs and, subsequently, impact homeostasis over time. Prolonged exposure (a state of chronic stress) can lead to the continuous release of corticosteroids with deleterious effects on the organ and immune systems. This includes the possible fatigue and dampening of the physiological stress response and suppressing immune system response to antigens. The prenatal and early postnatal development of the HPA axis can be affected by the presence of maternal stress and circulating corticosteroids which have long lasting effects on the endocrine, CNS and other systems (McEwen, 2019). No generalised endocrine profile exists for chronically stressed wild animals (Dickens and Romero, 2013), highlighting the complexity of measuring and interpreting chronic stress responses. We argue that there is also little evidence for a current endocrine profile of chronically stressed dogs.

Behavioural changes provide visible evidence of physiological responses to challenges. The SNS "fight or flight" response manifests in observable canine behaviours including activity level changes, facial and postural adjustments (e.g., eyes, mouth, ears). In addition, vocalisations and subtle indicators, like lip licking, panting, yawning, shake-offs, etc. can occur (Beerda et al., 1997; Beerda et al.,1999; Grigg et al., 2021).

**Beyond cortisol**

While physiological stress responses are well-documented across species (Hochachka and Somero, 2002), dog-specific evidence remains limited, often extrapolated from human and rodent studies. Recent decades have seen an increased focus on dog welfare (e.g., Beerda et al., 1999; Protopopova 2016; Edwards et al., 2019), but research has primarily measured HPA and SNS activation through cortisol and heart rate. This narrow focus potentially overlooks the complex interactions between the different physiological systems (Oken et al., 2015). Canine science often misinterprets stress responses by oversimplifying their relationship with glucocorticoid levels (MacDougall-Shackleton et al., 2019). Treating cortisol levels in dogs as a simple indicator of stress response state or welfare (low = good, high = bad) can lead to incorrect conclusions that can be harmful to dogs. Researchers may conclude an experience (e.g., novel environment, human interaction) is not aversive to dogs when, in fact, the experimental protocol has not adequately evaluated how dogs have responded. Multiple researchers have emphasized that glucocorticoids alone cannot adequately characterise vertebrate stress response (e.g., Breuner et al., 2013; Romero and Beattie, 2022). The increasing tendency to label cortisol (or other biomarkers) as "stress hormones" is inaccurate, further oversimplifying its complex role (Dickens and Romero, 2013; MacDougall-Shackleton et al., 2019).

Glucocorticoids serve multiple physiological functions beyond stress response (Broom, 2017; Vera et al., 2017; Gormally and Romero, 2020), including coordinating metabolic rates and daily activities, and regulating sleep patterns (McEwen, 2019). Cortisol serves age-specific functions across the lifespan of an individual: neonatal amygdala development, adolescent growth promotion, and potentially accelerating aging when found in excess (McEwen, 2019). Stressor-induced cortisol levels do not create homeostatic challenge (i.e. stress), they help minimise stress responses (MacDougall-Shackleton et al., 2019), and normally fluctuate given a certain environment or circadian cues (McEwen, 2019). There are some common misconceptions regarding cortisol concentration that should be clarified. Namely, that acute high cortisol concentrations equal an animal in distress when it really implies that the animal is appropriately coping with the stressor (Romero and Beattie, 2022). In turn, that low acute cortisol concentrations depict a healthy animal, when this could mean an animal that has stopped appropriately responding to its environment due to persistent stressors causing HPA axis fatigue and burnout (Romero and Beattie, 2022). Consistent elevation in cortisol doesn't necessarily indicate compromised welfare, since cortisol alone is not a good predictor of stress response state (Romero and Beattie, 2022).

When investigating physiological stress responses, study design must consider and report individual dog characteristics (genetics, personality, physical traits like weight), as well as the duration of stimulus exposure (e.g., acute, chronic, or unknown). The process of measurement itself can lead to spurious and invalid results. Collection method invasiveness ranges from non-invasive (e.g., collecting a post-excretion faecal sample), to mildly invasive (e.g., collection of hair or saliva), to more invasive (e.g., blood draw), to highly invasive (e.g., organ biopsy). Dogs' individual histories, any early-life trauma, training, or body sensitivity will affect their response to sampling methods. A key consideration includes the time delay between handling and physiological changes. Further replication is indicated to confirm if practices, such as blood samples taken within four minutes by an experienced person (as stated in Hennessy et al., 1997), avoid reflecting the physiological changes induced by handling. Pre-sampling training and habituation should also be considered. Study design using pre- and post-test responses, including a return to baseline within a single animal may alleviate some of these challenges.

Given domestic dogs' exceptional morphological and physiological diversity (Bryce et al., 2021), dog age and body weight variability (i.e., reference ranges) for each stress biomarker should be established and reported. Ideally, these should be considered prior to establishing the variability of these parameters in response to acute and chronic stressors. Additional considerations should include sex and reproductive status of the dog at the time that the biomarker was taken (refer Cobb et al., 2016). A clear understanding of the biomarkers' properties, circadian cycles, allostasis (refer Lu et al., 2021), and interpretation is also necessary. In the following section of this discussion, we explore indicators of the physiological stress response commonly used in canine science, as well as the opportunities for future biomarker development and interpretation.

**Physiological Indicators of the Stress Response in Dogs— A Summary of the Evidence.**

*Cortisol*

Circulating cortisol can be measured in multiple biological samples: blood, saliva, urine, faeces, and hair (Beerda et al., 1996; Schatz and Palme, 2001; van der Laan et al., 2022; Kooriyama and Ogata, 2021). Each sample type reflects different timeframes, from immediate circulating levels in blood to longer-term (weeks or months) in hair. The distribution of cortisol in these different modalities may also be affected by the presence of corticosteroid-binding globulin. Cortisol measurement in all sample types has been widely used as an indicator of stress response in canine research, though it is often misinterpreted due to a lack of contextual reporting and understanding (Broom, 2017). In a meta-analysis of data from 61 peer-reviewed studies using salivary cortisol, Cobb and colleagues (2016) found a significant effect on cortisol levels for sex and neuter status, age, environment, and sample collection media. No significant effects of dog breed, body weight, dog type, or coat colour were noted. Substantial variability between and among individuals complicates interpretation, makes it difficult to compare findings from different studies, and limits population-level conclusions from cortisol measurements alone. More recent work in dogs determined that dogs with known adverse early life backgrounds had variable salivary cortisol responses compared to other dogs in some situations (Buttner and Strasser, 2022; Buttner et al., 2023). These findings reiterate the significance of testing environment and attachment to people present on observed stress responses. Collection methods, subsequent sample processing, and materials also play a role in salivary cortisol measurement (Dreschel and Granger, 2009; Chmelíková et al., 2020). As with any biomarker, the potential impact from the sample collection experience must be considered and controlled for in the study design.

*Heart Rate and Heart Rate Variability*
Heart rate (HR) and heart rate variability (HRV) serve as proxies for SNS response in dogs (Amaya et al., 2020). Sympathetic activity increases HR, while parasympathetic activity decreases it (Bidoli et al., 2022). Excitement or anxiety activates the SNS and, in turn, increases catecholamine release from the adrenal medulla and output of norepinephrine from adrenergic nerve endings (Höglund et al., 2012). The root mean square of successive differences of inter-beat intervals (RMSSD) provides a measure of HRV (parasympathetic activity) through beat-to-beat variations (Amaya et al., 2020).

Research shows stressors typically increase HR and decrease in RMSSD in dogs (Katayama et al., 2016; Zupan et al., 2016). When approached by a threatening stranger, dogs showed increased HR regardless of the owner's presence, while another study showed HRV decreased in dogs only when their owner was absent (Gacsi et al., 2013). Similar HR changes occur during anxiety and fear responses in the Ainsworth's Strange Situation (Palestrini et al., 2005; Lu et al., 2021). Environmental studies have shown mixed results, with shelter dogs displaying unchanged HRV but increased rest behaviour during music exposure (Amaya et al., 2020). In animal-assisted education settings, dogs deemed "problematic" (behaviourally challenging) showed a significant increase in HR compared with their "innocuous" (not behaviourally challenging) counterparts (Bidoli et al., 2022). While HR and HRV can indicate stress states, the relationship between behavioural and HRV indicators remains debated (see Maros et al., (2008) and Bergamasco et al., 2010, for example). For a review on this topic on farm animals, please see von Borell et al., (2007). Clinical applications for HR and HRV in dogs are reviewed in Kartashova et al. (2021).

Understanding HR and HRV responses requires consideration of age and body weight effects. Hezzell and colleagues (2013) found HR to be negatively correlated with body weight in healthy dogs, though age and HR demonstrate a more complicated relationship with changes across the dog's lifespan. Some breeds, namely, Border Collie, Golden Retriever, Labrador Retriever and Springer Spaniel, demonstrated significantly lower HR; whereas, Cavalier King Charles Spaniels, Staffordshire Bull Terriers and Yorkshire Terriers had significantly higher HR (Hezzel et al., 2013). Most of the research on dogs of different body weights and ages does not seem to report on, nor correct for, these variables.

*Oxytocin*

Oxytocin, a peptide hormone produced in the hypothalamus and released by the posterior pituitary gland (Onaka, 2004), regulates social behaviour, attachment, anxiety, fear and stress response (Kikusui et al., 2019). It moderates stress responses by inhibiting ACTH secretion, thereby decreasing cortisol production (Suh et al., 1986; Li et al., 2019). Oxytocin is a hormone of interest in human-animal relationships, (Beetz et al., 2012; Carter and Porges, 2016; MacLean and Hare, 2015). Both species demonstrate an increase in oxytocin (evident in urine, blood and saliva) following positive social interactions (Nagasawa et al., 2009; 2015; MacLean et al., 2017). Oxytocin increases in urine of dogs following eating or exercise, suggesting this hormone may serve as an indicator of positive emotions (Mitsui et al., 2011). When administered as intranasal spray, oxytocin enhanced social play (Romero et al., 2015), improved dogs' understanding of human social cues (Oliva et al., 2015), and strengthened both intraspecific and interspecific bonds between dogs and humans (Romero et al., 2014). Intranasal spray of oxytocin also decreases HR and increases HRV in dogs, demonstrating the inter-connectedness between oxytocin and the SNS (Kris et al., 2014). Research on oxytocin production in dogs has not thoroughly examined effects of body weight or reproductive status, though limited evidence suggests age and sex influences (Mengoli et al., 2021). Breed differences exist in response to exogenous oxytocin (Kovács et al., 2016). This hormone can be measured in urine, blood and saliva in domestic dogs (MacLean et al., 2018). While oxytocin is measured peripherally, the action is at the brain level. Current research is focused on the transport of oxytocin across the blood-brain barrier (Higashida et al., 2024). The presence of these receptors likely affects the individual response to peripheral and exogenous oxytocin levels.

*Temperature*

Infrared thermography (IR) offers a non-invasive method to measure temperature changes as an indicator of stress responses in dogs. This technique detects surface temperature and micro-circulation variations, providing a visual representation of body temperature (Speakman and Ward, 1998) and insights into both acute and chronic stress states in animals (Stewart, et al., 2005; 2007). HPA axis activation during stress responses increases blood flow, changing an animal's heat production and loss (Bouwknecht et al., 2007). A positive correlation between core body temperature and stress response state is well-established (i.e. Bouwknecht et al., 2007). Eye temperature determined by IR strongly correlates with autonomic nervous system (ANS) activity (Travain et al., 2015).

Most vertebrates regulate internal body temperature (Tb) through behavioural and physiological mechanisms linked to metabolism (Zhao et al., 2022). Internal body temperature can increase due to a state of stress, thus termed "thermal stress". Body size significantly influences this regulation with larger animals having lower mass-specific metabolic rates than smaller ones (McNab 1997). This relationship reflects adaptations to heat dissipation based on the animal's surface area:volume ratio (SA:V). Smaller animals, having larger SA:V, exhibit higher heat production rates and mass-specific metabolic rates (Schmidt-Nielsen 1984). While large endothermic animals primarily generate heat through cellular metabolism, with a small contribution from thermogenesis, smaller ones lose heat at faster rates than larger ones. Thus, smaller endotherms should be able to orient their physiology towards heat production rather than the generation of adenosine triphosphate (ATP), the primary carrier of energy in cells, used for storing and transferring energy, when needed (Abreu-Vieira et al., 2015). The 44-fold variation in domestic dog body sizes creates significant differences in thermal regulation across breeds (Jimenez, 2016). Heat limitation particularly affects smaller breeds due to their inherently high metabolic heat production (Jimenez, 2016). This will also limit the thermal stress response of dogs. Body mass shows positive correlation with internal temperature and heat dissipation rates in dogs across seasons, while negatively correlating with mass-specific metabolism (Jimenez et al., 2023). Network analyses confirm body mass as a central factor in dogs' thermal regulation. While breed significantly effects thermoregulation, age and sex do not show notable influence (Jimenez et al., 2023).

IR studies reveal increased eye temperature in domestic dogs during stressful situations like veterinary visits, with peak core temperatures occurring during clinical examination when dogs exhibit frozen behaviour (Travain et al., 2015). This indicates that temperature changes reflect stress response rather than physical activity (Travain et al., 2015). Comprehensive coverage of IR principles and applications in dog welfare science is available in Travain and Valsecchi (2021). For a review on the usefulness of IR in a clinical setting, please see Casas-Alvarado et al. (2020) and Mota-Rojas et al. (2022).

*Salivary Alpha-amylase*

Salivary alpha-amylase (sAA) serves as an SNS indicator for psychosocial stress in humans, responding to social defeat, confrontations, and trauma (Lu, et al., 2021; Nater et al., 2006). However, its role in dogs remains uncertain. While α-amylase mRNA transcripts appear absent in the canine parotid glands, they exist in other tissue, including pancreas, liver, intestines, and fallopian tube (Mocharla et al., 1990). Though detectable at low concentrations in canine saliva (Conteras-Aquilar et al., 2017), sAA's stress-related changes lack investigation regarding age, weight, sex, or reproductive status effects (Kooriyama and Ogata, 2021).

*Melatonin*

Melatonin synchronises dogs' internal biological clock with the day-night cycles, with its concentration (in blood) showing natural fluctuations (Zan et al., 2013; Chojnowska et al., 2021). Beyond circadian regulation, melatonin influences sleep-wake cycles and responds to environmental cues including daylight and temperature (Saarela and Reiter, 1994). It also serves as an antioxidant in normal biological processes, like oxidative stress, where it helps to maintain balance between free radicals and antioxidants to prevent cellular damage and disease (Konturek et al., 2007). Salivary melatonin levels don't reliably reflect pineal gland production, making blood the preferred mode of measurement (Kennaway, 2020). In dogs, blood melatonin follows circadian patterns like humans, peaking at night and reaching lowest levels at dawn (Zan et al., 2013), with seasonal variation showing higher levels during thermally stressful (winter) months (Dunlap et al., 2007). While melatonin shows promise for understanding and improving mental states related to anxiety, aging, and sleep quality in dogs, research has yet to explore relationships between melatonin and factors such as age, body weight, sex, and reproductive status.

*Chromogranin A*

Chromogranin A (CgA) and its epitopes, catestatin and vasostatin, can be measured in both blood and saliva, remain stable across age, gender, breed, and time of day in healthy dogs (Srithunyarat et al., 2017). While blood levels show no significant stress-related changes, salivary catestatin increases significantly when dogs encounter stressful situations like pain, unfamiliar environments or sampling methods (Srithunyarat et al., 2018). CgA and its epitopes could be valuable tools for evaluating the effectiveness of stress-reduction interventions. However, comprehensive research on age and body weight effects remains limited (Kooriyama and Ogata, 2021).

*Arginine Vasopressin*

Arginine Vasopressin (AVP) levels correlate with specific behavioural patterns in dogs: higher plasma AVP has been associated with increased aggression towards other dogs (MacLean et al., 2017b), while salivary AVP increased in dogs lacking human interaction during human-animal interaction social studies (MacLean et al., 2017). Though AVP responds to acute stress (Kooriyama and Ogata (2021) and could provide insights into both stress states and aggressive predispositions, research has not yet examined effects of age, body weight, sex or reproductive status.

*Serotonin*

Serotonin influences multiple affective states and stress-related behaviours, including aggression, impulsivity, and reaction to pain (Berger et al., 2009). This neurotransmitter is commonly measured in blood, with peripheral serotonin representing a distinct pool from central serotonin due to the blood-brain barrier and playing an important role in regulating metabolic homeostasis (El-Merahbi et al., 2015) .  Lower blood serotonin levels in dogs correlate with reduced human sociability and increased aggression in dogs (Alberghina et al., 2017). While studies have investigated age and sex effects (Alberghina et al., 2017; 2019) and preliminarily explored body weight relationships (Bochiș, et al., (2022), sample sizes remain small and reproductive status effects unexplored (Ogi and Gazzano 2023).

*Microbiome/Microbiota Relationships*

Research across species demonstrate important multidirectional interactions between brain, gut microbiota, and stress response, including relationships with anxiety behaviours, early life stress effects on the adult microbiota, and microbial influence on HPA axis development (Dinan and Cryan 2012). Limited canine studies show variable results (Venable et al., 2016, Perry et al., 2017, Cannas et al., 2021), though one larger working dog study found microbiome markers predicted several relevant working dog characteristics, including motivation, sociability, and gastrointestinal issues, but not stress responses (Craddock et al., 2022). While microbiota markers hold promise as indicators of acute or long-term stress responses, specific recommendations do not currently exist.

*Immune Response*

Secretory Immunoglobulin A

Immunoglobulin A (IgA) is a group of antibodies found in the mucous membranes and its level in saliva has been shown respond to changes in mood in humans and both chronic and acute stress response states in dogs (Kooriyama and Ogata, 2021). Studies have found that sIgA levels in dogs tend to decrease following both acute and long-term stress, suggesting it could be used to measure stress responses. However, sIgA exhibits diurnal variation and may respond differently in puppies versus adult dogs, so further research is needed to fully validate its use as a reliable indicator of canine mental states and welfare across different contexts, time of day, and age groups (Skandakumar, et al., 1995; Kikkawa, et al., 2003; Svobodá, et al., 2014). Salivary IgA measurements require careful interpretation, considering factors such as saliva flow rate, dental disease, age (especially during tooth eruption), sex, reproductive status, and potential sample contamination (e.g., faecal, blood, etc.).

White Blood Cell Distribution

Leukocytosis (increased white blood cell numbers in blood) reflects sympathetic nervous system and/or HPA axis activation. The veterinary clinical pathology evidence refers to the response of white blood cells to epinephrine release as a "physiologic leukogram" (McCourt and Rizzi, 2022). This rapid, transient response involves neutrophil demargination from tissue to circulation, lymphocyte mobilisation from the thoracic duct and blockage of lymphocytes entering the lymph nodes, along with potential increases in monocytes and eosinophils (Stockham et al., 2003). The resulting broad increase in white blood cell counts in response to epinephrine, provides a measurable indicator of acute stress response. Increased red blood cell numbers (erythrocytosis) can also occur following epinephrine release due to splenic contraction (Stockham et al., 2003). In contrast, a "stress leukogram" reflects white blood cell count changes from endogenous or exogenous glucocorticoids (McCourt and Rizzi, 2022). Like the response to epinephrine, corticosteroid release triggers neutrophil movement from tissues to circulation and bone marrow release, (causing neutrophilia, and possible monocytosis), corticosteroid response differs by redistributing lymphocytes to bone marrow and lymphatic tissue, reducing blood lymphocyte count (Stockham et al., 2003). Eosinopenia can also occur through tissue repartitioning and decreased bone marrow release (McCourt and Rizzi, 2022). These changes – neutrophilia, monocytosis, lymphopenia, and eosinopenia – emerge within hours of a challenging experience and may persist for days, depending on corticosteroid exposure duration (McCourt and Rizzi, 2022). Neutrophil/Lymphocyte ratios serve as HPA axis stress response indicators across species (Davis et al., 2008), including studies of transportation-related stress in dogs (Herbel et al., 2020). However, interpretation requires careful consideration of sampling-induced stress effects from restraint and needle puncture, and the difficulty of distinguishing stress-related leukocyte changes from those caused by inflammation and infection.

*State of the evidence*

While research to date confirms various physiological indicators relate to stress response in dogs, most studies inadequately account for crucial variables like body weight, age, sex and reproductive status that influence the underlying physiological processes. Common indicators such as cortisol show significant individual variability, limiting their reliability as standalone measures. Advancing canine welfare science requires establishing standardised protocols and reference ranges that consider individual and demographic variation for these indicators. Several promising new biomarkers from other fields of research may also merit investigation in dogs to expand our measurement toolkit.

*Table 1. Summary of physiological measures.*

| Measure | Mode of Measurement | Acute or Chronic | References |
|---|---|---|---|
| Heart rhythm (HR, HRV) | Pulse, Stethoscope, ECG, HR monitor | Acute | Gacsi et al., 2013; Katayama et al., 2016; Zupan et al., 2016; Amaya et al., 2020 |
| Oxytocin | Blood, saliva, urine | Acute | Nagasawa et al., 2009; 2015; MacLean et al., 2017 |
| Cortisol | Plasma, saliva, urine, faeces, hair | Acute | Beerda et al., 1996; Cobb et al., 2016; Chmelíková et al., 2020; Schatz & Palme 2001; van der Laan et al., 2022 |
| Body temperature | IR thermography, thermometer (rectal, ear, microchip) | Acute, Chronic | Travain et al., 2015; Travain & Valsecchi, 2021 |
| Salivary Alpha-amylase (sAA) | Saliva | Acute | Kooriyama & Ogata, 2021 |
| Melatonin | Blood | Chronic | Dunlap et al., 2007; Zan et al., 2013; Kennaway, 2020 |
| Chromogranin A (CgA) | Blood, saliva | Acute | Srithunyarat et al., 2018; Kooriyama and Ogata 2021 |
| Arginine vasopressin (AVP) | Blood, saliva | Acute | MacLean et al., 2017b; Kooriyama and Ogata 2021 |
| Serotonin | Blood | Acute, Chronic | Alberghina et al., 2017; 2019 |
| Secretory IgA | Saliva | Acute, Chronic | Skandakumar et al., 1995; Kikkawa et al., 2003; Svobodová et al., 2014; Chojnowska et al., 2021; Kooriyama & Ogata, 2021 |
| White blood cell amounts and ratios | Blood | Acute, Chronic | McCourt & Rizzi, 2022, Stockham et al., 2003 |
| Microbiome | Faeces | Acute, Chronic | Craddock et al., 2022 |
| Lysozyme | Saliva | Acute in humans | Chojnowska et al., 2021 |
| Catecholamines (epinephrine and norepinephrine) | Blood | Acute | Not measured yet in dogs |
| Pro-inflammatory cytokines | Blood | Acute and Chronic | Jiménez and Strasser 2024 |

| Telomeres | Blood, mucosal tissue (cheek swab) | Chronic | Not measured yet in dogs |
| --- | --- | --- | --- |
| Oxidative stress measures | Blood | Acute, Chronic | Jiménez and Strasser 2024 |
| Osteocalcin | Blood | Acute | Not measured yet in dogs |
| CNS measures | MRI, EEG | | Not yet measured in relation to welfare |

**Opportunities for future research**

Several physiological indicators show promise for canine welfare assessment but remain unexplored. A brief description of these markers and what is known about them thus far in the canine or in other species follows.

*Lysozyme*

Lysozyme, a protein produced by immune cells (monocytes and macrophages), and present in saliva, provides innate defence against pathogens (Chojnowska et al., 2021). While human studies link it to acute stress response, its potential as a canine welfare indicator has not been explored.

*Catecholamines: Epinephrine and Norepinephrine*

Catecholamines are key biomarkers for ANS activity. Epinephrine and norepinephrine are mainly synthesised by the central noradrenergic neurons and are released mainly from the adrenal medulla into the bloodstream to mount the sympathetic-dominated "fight-or-flight" response (Segerstrom and Miller, 2004). Many surrogate markers of ANS activity (e.g. HR, respiratory rate, blood pressure, etc.) have been proposed. Despite their importance, research on these compounds in dogs remains limited, partly due to sampling challenges. As with other markers, the collection of blood for sampling could cause spurious results.

*Pro-inflammatory Cytokines*

Pro-inflammatory cytokines offer another promising avenue, particularly given the immune system's age-related changes in dogs. Aging affects T-cell immunity and immunoglobulin levels (Day, 2010), with studies showing decreased mitogen stimulation and changes in white blood cell populations (Greeley et al., 1996; Strasser et al., 1993). However, current research lacks comparisons between breeds and consideration of size-specific aging effects.

While chronic inflammation underlies many age-related diseases in humans, including cardiovascular disease, arthritis, and cancer (Chung et al., 2009; Davizon-Castillo et al., 2019) involving multiple cytokines and molecular pathways (Franceschi and Campisi, 2014), only limited data exists for dogs. Disease progression seems to be a causative factor linked to the inflammatory process (Chung et al., 2006). Humans demonstrate an increase in interleukin (IL)-6 and tumour necrosis factor α (TNF-α) levels with increased age and a decrease in insulin growth factor (IGF)-1 levels with increasing age (Xia et al., 2016). Proinflammatory cytokines IL-1β, TNF-α, and IL-6 in humans act both in paracrine and autocrine manners by developing insulin resistance, and they can interfere with insulin signalling, lipid and protein synthesis, potentially inducing metabolic disorders, including diabetes (Lontchi-Yimagou et al., 2013; Sung et al., 2018). One study found that increases in age demonstrated increased IL-6 concentrations in dog serum, but no changes with age in IL-1β, TNF-α (Jiménez, 2023). Another study found that shelter dogs, when compared with client-owned dogs, showed increased IL-1β levels, potentially leaving the shelter-dog populations at higher risk for zoonotic diseases (Jimenez and Strasser 2024). Human studies link chronic stress experiences to increased cytokine levels (Gouin et al., 2012; Lennartsson et al., 2016), suggesting potential applications in canine welfare assessment.

*Telomere length*

Telomere attrition shows promise as a potential biomarker for assessing cumulative stress and welfare in dogs. Telomeres are protective caps at the end of chromosomes that are linked to genomic stability. As cells divide with increasing organismal age, or in times of increased oxidative stress, these caps get shorter (Richter and von Zglinicki, 2007). Chronic stress experience has been associated with decreases in telomere length in humans and other animals. This work has been reviewed by Houben et al., (2008) and Mathur et al., (2016), with research suggesting that telomere length can provide a molecular measure of an animal's lifetime experiences, with negative experiences accelerating telomere shortening and positive interventions potentially mitigating or even reversing this attrition. There are no studies looking at stress and telomere length in domestic dogs, but others have established a link between telomere dynamics and the lifespan of different breeds of dogs without consideration for sex or reproductive status (Fick et al., 2012). Measuring telomere length could offer a valuable tool for evaluating the long-term impacts of various husbandry practices, environmental conditions, and life events on canine mental states and overall welfare between different populations of dogs.

*Oxidative stress*

Mitochondria structure and function play a central role in organismal homeostasis (Havird et al., 2019; Hood, et al., 2018; Calhoon, et al., 2014), acting as both targets and mediators of stress responses (Picard and McEwen, 2018). During "fight or flight" responses, they meet increased ATP demands (Manoli, et al., 2007), but approximately 5% of oxygen processed becomes a reactive oxygen species (ROS) (Hulbert et al., 2007). While low ROS levels serve as important signalling molecules for gene regulation, cell signalling, and cell processes (apoptosis) (Sohal and Orr, 2012; Dowling and Simmons 2009; Monaghan et al. 2009), high levels can damage lipids, proteins and even DNA (Finkel and Holbrook, 2000; Monaghan et al., 2009).

Oxidative stress reflects the balance between pro-oxidants from aerobic metabolism and protective antioxidants (Halliwell and Chirico, 1993; Ayala et al., 2014). Oxidative damage accrues when ROS production overwhelms the antioxidant system (Sohal and Orr, 2012; Dowling and Simmons, 2009; Monaghan et al., 2009). Oxidative damage can happen to many biologically relevant molecules, such as proteins, DNA and lipids (Hulbert et al., 2007). Lipids are among the molecules most affected, and two of the most prevalent pro-oxidants that can initiate damage to lipid membranes are hydroxyl radicals (OH•) and hydroperoxyl radicals (OOH•) (Ayala et al., 2014). The process of lipid peroxidation (LPO) continues unabated until the propagation of damage is halted by an antioxidant molecule (Halliwell and Chirico, 1993; Ayala et al., 2014). Enzymatic antioxidants, such as glutathione peroxidase (GPx), superoxide dismutase (SOD) and catalase (CAT), function by catalysing the oxidation of less biologically insulting molecules. Other antioxidant molecules, such as vitamin E and

C, act as chain-breaking antioxidants; they scavenge for ROS, remove them once they are formed, and further halt the propagation of peroxidation (Halliwell and Chirico, 1993). The following consideration is of utmost importance: The relationship between metabolism and ROS production is not straightforward, though many assume that an increase in oxygen consumption (i.e. increase in metabolic rate) should yield an increase in ROS production (Hou et al., 2020). Oxygen consumption can either be coupled with ATP production or heat depending on whether the ATP synthase or mitochondrial uncoupling proteins are driving respiration (Divakaruni and Brand, 2011; Hou et al., 2020), thus, it should never be assumed that this process is linear.

To accurately measure oxidative stress within an organism, the "damage" side and the antioxidant side should be measured together. That is, at least one measure of oxidative damage should be considered with at least one measurement of antioxidant capacity. Additionally, accurate estimations of whole-animal oxidative stress should be done in separated blood, as blood is a metabolic reservoir for these types of molecules (Jimenez and Downs, 2020) or in metabolically active tissues. Saliva and urine are not good sources for measuring oxidative stress in animals, and do not demonstrate a whole-animal perspective for the process. The connection between glucocorticoid secretion and OS increases has been documented in dogs (Ferreira et al., 2014); preliminary work has suggested OS could be used with dogs responding to different environments (e.g., shelters per Passantino, 2014). Age, body weight and breed considerations to oxidative stress have been measured in domestic dogs (Jimenez and Downs, 2020; Jiménez & Strasser, 2024), sex and reproductive status have also been considered using primary fibroblast cells (Jimenez et al., 2018; 2020).

*Osteocalcin*

Osteocalcin, an osteoblast-produced protein abundant in bone matrix, influences glucose metabolism and fat mass in metabolic homeostasis (Lieben et al., 2009). Berger and colleagues (2019) review the role of osteocalcin in mediation of the acute stress response. Exposure to a variety of physical and emotional stressors in rodents and humans leads to a rapid increase in circulating bioactive osteocalcin. There is evidence that this release leads to a suppression of the parasympathetic nervous system, allowing the acute stress response to proceed (Berger et al., 2019). Researchers have reported different patterns of change in serum osteocalcin and salivary cortisol in humans exposed to low- and high-intensity physical threat scenarios (Vít et al., 2023). Given its presence across vertebrates, osteocalcin warrants investigation as a canine stress biomarker, though factors like age, activity, nutrition, body condition, and other factors will likely play a role in its release and subsequent interpretation.

**How could we better represent animal experiences?**

With the increase in the use of artificial intelligence and newer technologies, there are great opportunities to combine and model the approach to stress and animal welfare measurement.

*Emerging technologies*

Recent technologies enable enhanced monitoring of canine behaviour and physiology through wearables (e.g., actigraphy for sleep-activity patterns, accelerometers for movement), ingestible sensors (e.g., microsensors for temperature, heart rate, respiratory rate), portable devices (e.g., electroencephalogram), and smart (sometimes responsive) environments. However, applying these technologies requires careful consideration of the potential impacts on dogs' experiences (Webber et al., 2022). For example, devices emitting ultrasound above human hearing range (20kHz) but within canine (45kHz) could induce anxiety or behavioural changes if dogs cannot avoid the sound (Grigg et al., 2021). Such unintended effects might compromise the very observations these technologies aim to capture. While these emerging tools offer exciting opportunities for welfare assessment and canine welfare science, limited dog-specific validation currently restricts their scientific value. Future advances may come from adaptation of precision livestock farming technologies to canine applications (e.g., Neethirajan, 2024).

Existing technologies also continue to expand research capabilities in canine welfare science. Non-invasive resting-state functional magnetic resonance imaging (RS-fMRI) allows investigation of brain-behaviour connectivity in awake, trained dogs (Berns et al., 2012). The ability to use these tools to predict the future success of working dogs based on trainability influences the welfare of those dogs going through the training process (Deshpande et al., 2024). Wider availability and decreasing costs relating to the use of this technology may open even more opportunities for its application in the study of canine welfare.

Artificial intelligence (AI) offers new opportunities for analysing canine welfare indicators. Recent studies demonstrate AI's potential for predicting working dog success through personality assessment (Amirhosseini, et al., 2024) and olfactory detection capabilities based on behavioural traits and environmental conditions (Eyre, et al., 2023). However, accurate prediction of working dog suitability remains challenging, likely due to epigenetics (Bray, et al., 2021). Dogs continue to change as individual genetic potential, personality, environment and learning experiences interact while dogs mature throughout their early years of life. Big data projects are becoming more common in canine science (e.g., ManyDogs; VetCompass; Generation Pup; The Dog Aging Project; DogslifeUK; etc.) and we expect these data sets will offer new insights into epigenetics and behaviour informatics for which AI will undoubtedly prove to be a valuable tool to advance canine welfare.

We hope to see such validation and wider incorporation of these technologies within research as canine experience indicators become routine inclusions within human-animal interaction research. For too long, the focus within the field of anthrozoology and even canine science, has been to maximise the benefit to people from dogs (MacLean et al., 2021). Funding bodies increasingly require canine-specific welfare measures, though comprehensive assessment demands substantial resources, including funding, equipment, personnel, time, and subject dogs.

*Behavioural factors in assessing canine welfare*

Although this discussion paper emphasises physiological indicators, behavioural factors are also important when assessing canine welfare. We know that many behavioural factors can influence how a dog copes in given situations. These include canine personality (Posluns, et al., 2017), attachment (Payne et al., 2015), temperamental traits like shy-boldness (Starling et al., 2013), and cognitive bias (Barnard, et al., 2018). Inclusion of observed behaviour (e.g.

McGreevy et al., 2012), qualitative behaviour analysis (e.g., Flint et al., 2024), and/or other survey-based profiling (e.g., Ley, et al., 2009) will serve to provide helpful information from which to contextualise and interpret physiological data.

*Research design and analysis*

No single experimental design suits all assessments of canine physiological indicators at this time. Study design depends on the focus - whether examining arousal, emotional valence, or acute versus chronic distress (Gormally and Romero, 2020; Dickens and Romero, 2013). Modern research increasingly incorporates dog agency, allowing subjects to opt-in or opt-out of participation (Horowitz, 2021; Littlewood et al., 2023), reflecting emerging concepts of canine consent and emotional needs (Jones, 2024). Environmental presentation order can also influence canine responses (Paul et al., 2023), adding another layer of complexity. Research design must balance multiple interacting factors: research questions, statistical approaches, project budget constraints, timelines, sample size and availability of participating dogs, and sampling invasiveness. These considerations inform our recommendations while acknowledging the need for continued discussion and methodological advancement in canine welfare science.

Key recommendations:
1. Avoid using salivary cortisol alone to infer canine affect and welfare.
2. Design research that records multiple indicators of the dog experience. Ideally, these will rely on a) minimally invasive sampling that is b) most relevant to the system being evaluated in c) the context of that system's role in responding to the experience being evaluated.
3. Include validated behavioural measures and individual characteristics (such as dogs' sex, age, breed, weight, personality, owner presence, etc.) in study design and reporting, acknowledging that physiological systems' interactions may yield varying interpretations. This comprehensive approach enables more meaningful cross-study comparisons and advances field-wide understanding.

**Animal welfare implications**

This discussion advances canine welfare assessment by highlighting traditional biomarker limitations and advocating for a comprehensive systems approach, paving the way for more accurate and comprehensive assessments of canine welfare. Emphasising the importance of individual factors and appropriate research design enables a better understanding of positive and negative affective states in dogs, which is crucial for promoting their wellbeing. While significant work remains to be done in exploring physiological indicators, we encourage further developments in this field and welcome debate. The recommendations provided can lead to improved practices, enhanced scientific understanding, and better outcomes for the welfare of dogs.

**Conclusion**

Canine welfare science requires multiple physiological indicators because welfare's complexity cannot be captured by any single measure. Converging evidence from various indicators (like cortisol, heart rate, immune markers and behaviour) validates that we are evaluating welfare rather than some other physiological process (like age) or external factor.

We advocate for practical, evidence-based approaches using non-invasive methods while acknowledging the challenges in evaluating physiological indicators across different situations. Continued refinement of practices and terminology will enhance our understanding of dog welfare and inform assessment standards. This comprehensive approach promises to advance both scientific knowledge and practical care, ultimately improving the lives of dogs.

## Acknowledgements

AGJ received a Colgate University Research Council Faculty Leave grant to complete this work. MLC was supported during this work by The Chaser Fellowship in Canine Welfare Science. The authors thank Dr Kat Littlewood (Massey University, New Zealand) and Dr Birte Neilsen for the conversations that informed this manuscript and the reviewers for their valuable feedback.

## References

Abreu-Vieira, G., Xiao, C., Gavrilova, O., & Reitman, M. L. (2015). Integration of body temperature into the analysis of energy expenditure in the mouse. *Molecular metabolism*, *4*(6), 461-470.

Alberghina, D., Rizzo, M., Piccione, G., Giannetto, C., & Panzera, M. (2017). An exploratory study about the association between serum serotonin concentrations and canine-human social interactions in shelter dogs (Canis familiaris). *Journal of Veterinary Behavior*, *18*, 96-101.

Alberghina, D., Tropia, E., Piccione, G., Giannetto, C., & Panzera, M. (2019). Serum serotonin (5-HT) in dogs (Canis familiaris): Preanalytical factors and analytical procedure for use of reference values in behavioral medicine. *Journal of Veterinary Behavior*, *32*, 72-75.

Amaya, V., Paterson, M. B., Descovich, K., & Phillips, C. J. (2020). Effects of olfactory and auditory enrichment on heart rate variability in shelter dogs. *Animals*, *10*(8), 1385.

Amirhosseini, M. H., Yadav, V., Serpell, J. A., Pettigrew, P., & Kain, P. (2024). An artificial intelligence approach to predicting personality types in dogs. *Scientific Reports*, *14*(1), 2404.

Ayala, A., Muñoz, M. F., & Argüelles, S. (2014). Lipid peroxidation: production, metabolism, and signaling mechanisms of malondialdehyde and 4-hydroxy-2-nonenal. *Oxidative medicine and cellular longevity*, *2014*.

Barnard, S., Wells, D. L., Milligan, A. D., Arnott, G., & Hepper, P. G. (2018). Personality traits affecting judgement bias task performance in dogs (Canis familiaris). *Scientific Reports*, *8*(1), 6660.

Beausoleil, N. J., Swanson, J. C., McKeegan, D. E. F., & Croney, C. C. (2023). Application of the five domains model to food chain management of animal welfare: opportunities and constraints. *Frontiers in Animal Science*, 4:1042733.

Beerda, B., Schilder, M.B.H., Janssen, N. Mol, J.A. (1996).The Use of Saliva Cortisol, Urinary Cortisol, and Catecholamine Measurements for a Noninvasive Assessment of Stress Responses in Dogs, *Hormones and Behavior, 30*(3), 272-279.

Beerda, B., Schilder, M. B., Bernadina, W., Van Hooff, J. A., De Vries, H. W., & Mol, J. A. (1999). Chronic stress in dogs subjected to social and spatial restriction. II. Hormonal and immunological responses. *Physiology & behavior*, *66*(2), 243-254.

Beetz, A., Uvnäs-Moberg, K., Julius, H., & Kotrschal, K. (2012). Psychosocial and psychophysiological effects of human-animal interactions: the possible role of oxytocin. *Frontiers in psychology*, *3*, 234.

Bergamasco, L., Osella, M. C., Savarino, P., Larosa, G., Ozella, L., Manassero, M., ... & Re, G. (2010). Heart rate variability and saliva cortisol assessment in shelter dog: Human–animal interaction effects. *Applied animal behaviour science*, *125*(1-2), 56-68.

Berger, J.M., Singh, P., Khrimian, L., Morgan, D.A., Chowdhury, S., Arteaga-Solis, E., Horvath, T.L., Domingos, A.I., Marsland, A.L., Yadav, V.K., Rahmouni, K., Gao, X-B., & Karsenty, G. (2019). Mediation of the acute stress response by the skeleton. *Cell metabolism, 30*, 890-902.

Berger, M., Gray, J. A., & Roth, B. L. (2009). The expanded biology of serotonin. *Annual review of medicine*, *60*, 355-366.

Berns, G. S., Brooks, A. M., & Spivak, M. (2012). Functional MRI in awake unrestrained dogs. *PloS one*, *7*(5), e38027.

Bidoli, E. M., Erhard, M. H., & Döring, D. (2022). Heart rate and heart rate variability in school dogs. *Applied Animal Behaviour Science*, *248*, 105574.

Bochiș, T. A., Imre, K., Marc, S., Vaduva, C., Florea, T., Dégi, J., ... & Ţibru, I. (2022). The Variation of Serotonin Values in Dogs in Different Environmental Conditions. *Veterinary Sciences*, *9*(10), 523.

Bouwknecht, J. A., Olivier, B., & Paylor, R. E. (2007). The stress-induced hyperthermia paradigm as a physiological animal model for anxiety: a review of pharmacological and genetic studies in the mouse. *Neuroscience & Biobehavioral Reviews*, *31*(1), 41-59.

Bray, E. E., Otto, C. M., Udell, M. A., Hall, N. J., Johnston, A. M., & MacLean, E. L. (2021). Enhancing the selection and performance of working dogs. *Frontiers in veterinary science*, *8*, 644431.

Breuner, C. W., Delehanty, B., & Boonstra, R. (2013). Evaluating stress in natural populations of vertebrates: total CORT is not good enough. *Functional Ecology*, *27*(1), 24-36.

Broom, D. (2017). Cortisol: often not the best indicator of stress and poor welfare. *Physiology News*, *107*, 30-32.

Browning, H., & Birch, J. (2022). Animal sentience. *Philosophy compass*, *17*(5), e12822.

Browning, H., & Veit, W. (2023). Studying animal feelings: Integrating sentience research and welfare science. *Journal of Consciousness Studies*, *30*(7-8), 196-222.

Bryce, C. M., Davis, M. S., Gompper, M. E., Hurt, A., Koster, J. M., Larson, G., ... & Jimenez, A. G. (2021). Biology's best friend: bridging disciplinary gaps to advance canine science. *Integrative and Comparative Biology*, *61*(1), 76-92.

Bouyoucos, I.A., Schoen, A.N., Wahl, R.C., & Anderson, W. G. (2021). Ancient fishes and the functional evolution of the corticosteroid stress response in vertebrates. *Comparative Biochemistry and Physiology Part A: Molecular & Integrative Physiology, 260*, 111024.

Buttner, A. P., & Strasser, R. (2022). Extreme life histories are associated with altered social behavior and cortisol levels in shelter dogs. *Applied Animal Behaviour Science*, *256*, 105693.

Buttner, A. P., Awalt, S. L., & Strasser, R. (2023). Early life adversity in dogs produces altered physiological and behavioral responses during a social stress-buffering paradigm. *Journal of the Experimental Analysis of Behavior*, 120(1), 6-20.

Caddiell, R.M.P., Cunningham, R.M, White, P.A., Lascelles, B.D.X. & Gruen, M.E. (2023). Pain sensitivity differs between dog breeds but not in the way veterinarians believe. *Frontiers in Pain Research, 4*,1165340. doi: 10.3389/fpain.2023.1165340

Calhoon, E. A., Jimenez, A. G., Harper, J. M., Jurkowitz, M. S., & Williams, J. B. (2014). Linkages between mitochondrial lipids and life history in temperate and tropical birds. *Physiological and Biochemical Zoology*, *87*(2), 265-275.

Cannas, S. , Tonini, B., Belà, B., Di Prinzio, R., Pignataro, G., Di Simone, D., & Gramenzi, A. Effect of a novel nutraceutical supplement (Relaxigen Pet dog) on the fecal microbiome and stress-related behaviors in dogs: A pilot study. Journal of Veterinary Behavior-Clinical Applications and Research. 42, 37-47. Doi:10.1016/j.jveb.2020.09.002

Carter, C. S., & Porges, S. W. (2016). Neural mechanisms underlying human-animal interaction: An evolutionary perspective. In L. S. Freund, S. McCune, L. Esposito, N. R. Gee, & P. McCardle (Eds.), *The social neuroscience of human-animal interaction* (pp. 89–105). American Psychological Association. https://doi.org/10.1037/14856-006

Casas-Alvarado, A., Mota-Rojas, D., Hernández-Ávalos, I., Mora-Medina, P., Olmos-Hernández, A., Verduzco-Mendoza, A., ... & Martínez-Burnes, J. (2020). Advances in infrared thermography: Surgical aspects, vascular changes, and pain monitoring in veterinary medicine. *Journal of Thermal Biology*, *92*, 102664.

Chmelíková, E., Bolechová, P., Chaloupková, H., Svobodová, I., Jovičić, M., & Sedmíková, M. (2020). Salivary cortisol as a marker of acute stress in dogs: a review. *Domestic animal endocrinology*, *72*, 106428.

Chojnowska, S., Ptaszyńska-Sarosiek, I., Kępka, A., Knaś, M., & Waszkiewicz, N. (2021). Salivary biomarkers of stress, anxiety and depression. *Journal of clinical medicine*, *10*(3), 517.

Chrousos, G.P. (1998). Stressors, stress, and neuroendocrine integration of the adaptive response: The 1997 Hans Selye Memorial Lecture. *Annals of the New York Academy of Sciences, 851*(1), 311- 335.

Chung, H. Y., Sung, B., Jung, K. J., Zou, Y., & Yu, B. P. (2006). The molecular inflammatory process in aging. *Antioxidants & redox signaling, 8*(3-4), 572-581.

Chung, H. Y., Cesari, M., Anton, S., Marzetti, E., Giovannini, S., Seo, A. Y., ... & Leeuwenburgh, C. (2009). Molecular inflammation: underpinnings of aging and age-related diseases. *Ageing research reviews, 8*(1), 18-30.

Cobb, M.L., Iskandarani, K., Chinchilli, V.M., & Dreschel, N.A. (2016). A systematic review and meta-analysis of salivary cortisol measurement in domestic canines. *Domestic Animal Endocrinology*. 57, 31-42.

Cobb, M. L., Otto, C. M., & Fine, A. H. (2021). The animal welfare science of working dogs: current perspectives on recent advances and future directions. *Frontiers in veterinary science*, *8*, 666898.

Craddock, H.A., Godneva, A., Rothschild, D., Motro, Y., Grinstein, D., Lotem-Michaeli, Y., Narkiss, T., Segal, E., & Moran-Gilad, J. (2022). Phenotypic correlates of the working dog microbiome. *npj Biofilms Microbiomes 8*, 66. https://doi-org.ezaccess.libraries.psu.edu/10.1038/s41522-022-00329-5

Davis, A.K., Maney, D.L. & Maerz, J.C. (2008). The use of leukocyte profiles to measure stress in vertebrates: A review for ecologists. *Functional Ecology , 22*(5), 760-772.

Davizon-Castillo, P., McMahon, B., Aguila, S., Bark, D., Ashworth, K., Allawzi, A., ... & Di Paola, J. (2019). TNF-α–driven inflammation and mitochondrial dysfunction define the platelet hyperreactivity of aging. *Blood, The Journal of the American Society of Hematology*, *134*(9), 727-740.

Day, M. J. (2010). Ageing, immunosenescence and inflammageing in the dog and cat. *Journal of Comparative Pathology*, *142*, S60-S69.

de Lahunta, A., Glass, E., & Kent, M. (2021). *7- Lower Motor Neuron: General Visceral Efferent System*, Editors: Alexander de Lahunta, Eric Glass, Marc Kent, de Lahunta's Veterinary Neuroanatomy and Clinical Neurology (Fifth Edition), W.B. Saunders, 203-229.

De Winkel, T., van der Steen, S., Enders-Slegers, M. J., Griffioen, R., Haverbeke, A., Groenewoud, D., & Hediger, K. (2024). Observational Behaviors and Emotions to Assess Welfare of Dogs A Systematic Review. *Journal of Veterinary Behavior, 72*, 1-17.

Deshpande, G.; Zhao, S., Waggoner, P., Beyers, R., Morrison, E., Huynh, N., Vodyanoy, V., Denney, T.S., Jr., & Katz, J.S. (2024). Two Separate Brain Networks for Predicting Trainability and Tracking Training-Related Plasticity in Working Dogs. *Animals, 14*, 1082.

Dickens, M. J., & Romero, L. M. (2013). A consensus endocrine profile for chronically stressed wild animals does not exist. *General and comparative endocrinology*, *191*, 177-189.

Dinan, T.G. & Cryan, J.F. (2012). Regulation of the stress response by the guy microbiota: Implications for psychoneuroendocrinology. *Psychoneuroendocrinology, 37*, 1369-1378.

Divakaruni, A. S., & Brand, M. D. (2011). The regulation and physiology of mitochondrial proton leak. *Physiology*, *26*(3), 192-205.

Dowling, D. K., & Simmons, L. W. (2009). Reactive oxygen species as universal constraints in life-history evolution. *Proceedings of the Royal Society B: Biological Sciences*, *276*(1663), 1737-1745.

Dreschel, N.A., & Granger, D.A. (2005). Physiological and behavioral reactivity to stress in thunderstorm-phobic dogs and their caregivers. *Applied Animal Behaviour Science, 95* (3-4), 153-168.

Dreschel, N.A. & Granger, D.A. (2009). Methods of collection for salivary cortisol measurement in dogs. *Hormones and Behavior, 55*, 163-168.

Dunlap, K. L., Reynolds, A. J., Tosini, G., Kerr, W. W., & Duffy, L. K. (2007). Seasonal and diurnal melatonin production in exercising sled dogs. *Comparative Biochemistry and Physiology Part A: Molecular & Integrative Physiology*, *147*(4), 863-867.

Edwards, P. T., Smith, B. P., McArthur, M. L., & Hazel, S. J. (2019). Fearful Fido: Investigating dog experience in the veterinary context in an effort to reduce distress. *Applied Animal Behaviour Science*, *213*, 14-25.

El-Merahbi, R., Löffler M., Mayer, A., & Sumara, G. (2015), The roles of peripheral serotonin in metabolic homeostasis, *FEBS Letters*, 589, doi: 10.1016/j.febslet.2015.05.054

Eyre, A. W., Zapata, I., Hare, E., Serpell, J. A., Otto, C. M., & Alvarez, C. E. (2023). Machine learning prediction and classification of behavioral selection in a canine olfactory detection program. *Scientific Reports*, *13*(1), 12489.

Ferreira, C. S., Vasconcellos, R. S., Pedreira, R. S., Silva, F. L., Sá, F. C., Kroll, F. S., ... & Carciofi, A. C. (2014). Alterations to oxidative stress markers in dogs after a short-term stress during transport. *Journal of nutritional science*, *3*, e27.

Fick, L. J., Fick, G. H., Li, Z., Cao, E., Bao, B., Heffelfinger, D., ... & Riabowol, K. (2012). Telomere length correlates with life span of dog breeds. *Cell reports*, *2*(6), 1530-1536.

Finkel, T., & Holbrook, N. J. (2000). Oxidants, oxidative stress and the biology of ageing. *Nature*, *408*(6809), 239-247.

Flint, H. E., Weller, J. E., Parry-Howells, N., Ellerby, Z. W., McKay, S. L., & King, T. (2024). Evaluation of indicators of acute emotional states in dogs. *Scientific Reports*, *14*(1), 6406.

Franceschi, C., & Campisi, J. (2014). Chronic inflammation (inflammaging) and its potential contribution to age-associated diseases. *Journals of Gerontology Series A: Biomedical Sciences and Medical Sciences*, 69(Suppl_1), S4-S9.

Gácsi, M., Maros, K., Sernkvist, S., Faragó, T., & Miklósi, Á. (2013). Human analogue safe haven effect of the owner: behavioural and heart rate response to stressful social stimuli in dogs. *PLoS One*, *8*(3), e58475.

Gormally, B. M., & Romero, L. M. (2020). What are you actually measuring? A review of techniques that integrate the stress response on distinct time-scales. *Functional Ecology*, *34*(10), 2030-2044.

Gouin, J. P., Glaser, R., Malarkey, W. B., Beversdorf, D., & Kiecolt-Glaser, J. (2012). Chronic stress, daily stressors, and circulating inflammatory markers. *Health Psychology*, *31*(2), 264.

Granger, D., Dreschel, N., & Shirtcliff, E. (2003). Developmental Psychoneuroimmunology: The Role of Cytokine Network Activation in the Epigenesis of Developmental Psychopathology. In D. Cicchetti & E. Walker (Eds.), *Neurodevelopmental Mechanisms in Psychopathology* (pp. 293-323). Cambridge: Cambridge University Press. doi:10.1017/CBO9780511546365.014

Greeley, E. H., Kealy, R. D., Ballam, J. M., Lawler, D. F., & Segre, M. (1996). The influence of age on the canine immune system. *Veterinary immunology and immunopathology*, 55(1-3), 1-10.

Grigg, E.K., Chou, J., Parker, E., Gatesy-Davis, A., Clarkson, S.T. & Hart, L.A. (2021). Stress-related behaviors in companion dogs exposed to common household noises, and owners' interpretations of their dogs' behaviors. *Frontiers in Veterinary Science, 8*.

Halliwell, B., & Chirico, S. (1993). Lipid peroxidation: its mechanism, measurement, and significance. *The American journal of clinical nutrition*, *57*(5), 715S-725S.

Hampton, J. O., Jones, B., & McGreevy, P. D. (2020). Social license and animal welfare: Developments from the past decade in Australia. *Animals*, *10*(12), 2237.

Havird, J. C., Weaver, R. J., Milani, L., Ghiselli, F., Greenway, R., Ramsey, A. J., ... & Hill, G. E. (2019). Beyond the powerhouse: integrating mitonuclear evolution, physiology, and theory in comparative biology. *Integrative and comparative biology*, *59*(4), 856-863.

Hennessy, M. B., Davis, H. N., Williams, M. T., Mellott, C., & Douglas, C. W. (1997). Plasma cortisol levels of dogs at a county animal shelter. *Physiology & behavior*, *62*(3), 485-490.

Herbel, J., Aurich, J., Gautier, C., Melchert, M., & Aurich, C. (2020). Stress response of beagle dogs to repeated short-distance road transport. *Animals, 10*(11):2114. doi: 10.3390/ani10112114.

Hezzell, M. J., Humm, K., Dennis, S. G., Agee, L., & Boswood, A. (2013). Relationships between heart rate and age, bodyweight and breed in 10,849 dogs. *Journal of Small Animal Practice*, *54*(6), 318-324.

Higashida, H., Oshima, Y., & Yamamoto, Y. (2024). Oxytocin transported from the blood across the blood-brain barrier by receptor for advanced glycation end-products (RAGE) affects brain function related to social behavior. *Peptides*, *178*, 171230.

Hochachka, P. W., & Somero, G. N. (2002). *Biochemical Adaptation: Mechanism and Process in Physiological Evolution*. Oxford University Press.

Höglund, K., Hanås, S., Carnabuci, C., Ljungvall, I., Tidholm, A., & Häggström, J. (2012). Blood pressure, heart rate, and urinary catecholamines in healthy dogs subjected to different clinical settings. *Journal of veterinary internal medicine*, *26*(6), 1300-1308.

Hood, W. R., Austad, S. N., Bize, P., Jimenez, A. G., Montooth, K. L., Schulte, P. M., ... & Salin, K. (2018). The mitochondrial contribution to animal performance, adaptation, and life-history variation. *Integrative and Comparative Biology*, *58*(3), 480-485.

Hou, C., Metcalfe, N. B., & Salin, K. (2021). Is mitochondrial reactive oxygen species production proportional to oxygen consumption? A theoretical consideration. *BioEssays*, 2000165.

Houben, J. M., Moonen, H. J., van Schooten, F. J., & Hageman, G. J. (2008). Telomere length assessment: biomarker of chronic oxidative stress?. *Free radical biology and medicine*, *44*(3), 235-246.

Hulbert, A. J., Pamplona, R., Buffenstein, R., & Buttemer, W. A. (2007). Life and death: metabolic rate, membrane composition, and life span of animals. *Physiological reviews*, *87*(4), 1175-1213.

Jimenez, A. G. (2016). Physiological underpinnings in life-history trade-offs in man's most popular selection experiment: the dog. *Journal of Comparative Physiology B*, *186*(7), 813-827.

Jimenez, A. G., Winward, J., Beattie, U., & Cipolli, W. (2018). Cellular metabolism and oxidative stress as a possible determinant for longevity in small breed and large breed dogs. *PLoS One*, *13*(4), e0195832.

Jimenez, A. G., Winward, J. D., Walsh, K. E., & Champagne, A. M. (2020). Effects of membrane fatty acid composition on cellular metabolism and oxidative stress in dermal fibroblasts from small and large breed dogs. *Journal of Experimental Biology*, *223*(12), jeb221804.

Jimenez, A. G., & Downs, C. J. (2020). Untangling life span and body mass discrepancies in canids: phylogenetic comparison of oxidative stress in blood from domestic dogs and wild canids. *American Journal of Physiology-Regulatory, Integrative and Comparative Physiology*.

Jimenez, A. G., Paul, K., Zafar, A., & Ay, A. (2023). Effect of different masses, ages, and coats on the thermoregulation of dogs before and after exercise across different seasons. *Veterinary Research Communications*, 1-15.

Jiménez, A. G. (2023). Inflammaging in domestic dogs: basal level concentrations of IL-6, IL-1β, and TNF-α in serum of healthy dogs of different body sizes and ages. *Biogerontology*, *24*(4), 593-602.

Jiménez, A. G., & Strasser, R. (2024). Effects of Adverse Life History on Oxidative Stress and Cytokine Concentration in Domestic Dogs. *Journal of Applied Animal Welfare Science*, 1-13.

Jones, E. (2024). *Constructing Canine Consent: Conceptualising and adopting a consent-focused relationship with dogs*. CRC Press.

Katayama, M., Kubo, T., Mogi, K., Ikeda, K., Nagasawa, M., & Kikusui, T. (2016). Heart rate variability predicts the emotional state in dogs. *Behavioural processes*, *128*, 108-112.

Kartashova, I. A., Ganina, K. K., Karelina, E. A., & Tarasov, S. A. (2021). How to evaluate and manage stress in dogs–A guide for veterinary specialist. *Applied Animal Behaviour Science*, *243*, 105458.

Kennaway, D. J. (2020). Measuring melatonin by immunoassay. *Journal of Pineal Research*, *69*(1), e12657.

Kikkawa, A., Uchida, Y., Nakade, T., & Taguchi, K. (2003). Salivary secretory IgA concentrations in beagle dogs. *Journal of Veterinary Medical Science*, *65*(6), 689-693.

Kikusui, T., Nagasawa, M., Nomoto, K., Kuse-Arata, S., & Mogi, K. (2019). Endocrine regulations in human–dog coexistence through domestication. *Trends in Endocrinology & Metabolism*, *30*(11), 793-806.

Konturek, S. J., Konturek, P. C., Brzozowska, I., Pawlik, M., Sliwowski, Z., Cześnikiewicz-Guzik, M., ... & Pawlik, W. W. (2007). Localization and biological activities of melatonin. *Journal of Physiology and Pharmacology*, *58*(3), 381-405.

Kooriyama, T., & Ogata, N. (2021). Salivary stress markers in dogs: Potential markers of acute stress. *Research in Veterinary Science*, *141*, 48-55.

Kovács, K., Kis, A., Pogány, Á., Koller, D., & Topál, J. (2016). Differential effects of oxytocin on social sensitivity in two distinct breeds of dogs (Canis familiaris). *Psychoneuroendocrinology*, *74*, 212-220.

Kis, A., Kanizsár, O., Gácsi, M., & Topál, J. (2014). Intranasally administered oxytocin decreases heart rate and increases heart rate variability in dogs. *Journal of Veterinary Behavior: Clinical Applications and Research*, *9*(6), e15.

Lawrence, A. B., Vigors, B., & Sandøe, P. (2019). What is so positive about positive animal welfare?—a critical review of the literature. *Animals*, *9*(10), 783.

Lennartsson, A. K., Theorell, T., Kushnir, M. M., & Jonsdottir, I. H. (2016). Changes in DHEA-s levels during the first year of treatment in patients with clinical burnout are related to health development. *Biological psychology*, *120*, 28-34.

Ley, J. M., Bennett, P. C., & Coleman, G. J. (2009). A refinement and validation of the Monash Canine Personality Questionnaire (MCPQ). *Applied Animal Behaviour Science*, *116*(2-4), 220-227.

Li, Y., Hassett, A. L., & Seng, J. S. (2019). Exploring the mutual regulation between oxytocin and cortisol as a marker of resilience. *Archives of psychiatric nursing*, *33*(2), 164-173.

Lieben, L., Callewaert, F. & Bouillon, R. (2009). Bone and metabolism: A complex crosstalk. *Hormone Research, 71*(suppl1), 134-138.

Littlewood, K. E., Heslop, M. V., & Cobb, M. L. (2023). The agency domain and behavioral interactions: assessing positive animal welfare using the Five Domains Model. *Frontiers in Veterinary Science*, *10*, 1284869.

Lontchi-Yimagou, E., Sobngwi, E., Matsha, T. E., & Kengne, A. P. (2013). Diabetes mellitus and inflammation. *Current diabetes reports*, *13*(3), 435-444.

Lu, S., Wei, F. & Li, G. (2021). The evolution of the concept of stress and the framework of the stress system. *Cell stress, 5*(6), 76-85.

Lush, J., & Ijichi, C. (2018). A preliminary investigation into personality and pain in dogs. *Journal of Veterinary Behavior*, *24*, 62-68.

MacDougall-Shackleton, S. A., Bonier, F., Romero, L. M., & Moore, I. T. (2019). Glucocorticoids and "stress" are not synonymous. *Integrative Organismal Biology*, *1*(1), obz017.

Mackenzie, C., Rogers, W., & Dodds, S. (2014). Introduction: what is vulnerability and why does it matter for moral theory?. In *Vulnerability: new essays in ethics and feminist philosophy* (pp. 1-29). Oxford University Press.

MacLean, E. L., & Hare, B. (2015). Dogs hijack the human bonding pathway. *Science*, *348*(6232), 280-281.

MacLean, E. L., Gesquiere, L. R., Gee, N. R., Levy, K., Martin, W. L., & Carter, C. S. (2017). Effects of affiliative human–animal interaction on dog salivary and plasma oxytocin and vasopressin. *Frontiers in psychology*, 1606.

MacLean, E. L., Gesquiere, L. R., Gruen, M. E., Sherman, B. L., Martin, W. L., & Carter, C. S. (2017b). Endogenous oxytocin, vasopressin, and aggression in domestic dogs. *Frontiers in psychology*, *8*, 1613.

MacLean, E. L., Gesquiere, L. R., Gee, N., Levy, K., Martin, W. L., & Carter, C. S. (2018). Validation of salivary oxytocin and vasopressin as biomarkers in domestic dogs. *Journal of neuroscience methods*, *293*, 67-76.

MacLean, E. L., Fine, A., Herzog, H., Strauss, E., & Cobb, M. L. (2021). The new era of canine science: reshaping our relationships with dogs. *Frontiers in Veterinary Science*, *8*, 675782.

Manoli, I., Alesci, S., Blackman, M. R., Su, Y. A., Rennert, O. M., & Chrousos, G. P. (2007). Mitochondria as key components of the stress response. *Trends in Endocrinology & Metabolism*, *18*(5), 190-198.

Maros, K., Dóka, A., & Miklósi, Á. (2008). Behavioural correlation of heart rate changes in family dogs. *Applied Animal Behaviour Science*, *109*(2-4), 329-341.

Mason, G., & Mendl, M. (1993). Why is there no simple way of measuring animal welfare?. *Animal Welfare*, *2*(4), 301-319.

Mason, G. J. (2023). Animal welfare research is fascinating, ethical, and useful—but how can it be more rigorous?. *BMC Biology*, *21*(1), 302.

Mathur, M.B., Epel, E., Kind, S., Desai, M., Parks, C.G. & Sandler, D.P. (2016). Perceived stress and telomere length: A systematic review, meta-analysis, and methodologic considerations for advancing the field. *Brain, Behavior, and Immunity, 54*, 158-169.

McCorry, L.K., (2007).Physiology of the autonomic nervous system. *American Journal of Pharmaceutical Education, 71*(4), 78.

McCourt, M.R. & Rizzi, T.E. (2022). Hematology of Dogs. In M.B. Brooks, K.E.Harr, D.M.Seelig, K.J. Wardrop & D.J. Weiss (Eds.), *Schalm's Veterinary Hematology* (pp. 971-981). Hoboken, NJ: John Wiley & Sons, Inc.

McEwen, B.S. (2001). From molecules to mind: stress, individual differences, and the social environment. *Annals of the New York Academy of Sciences, 935*, 42-49.

McEwen, B. S. (2019). What is the confusion with cortisol?. *Chronic Stress*, *3*, 2470547019833647.

McGreevy, P. D., Starling, M., Branson, N. J., Cobb, M. L., & Calnon, D. (2012). An overview of the dog–human dyad and ethograms within it. *Journal of Veterinary Behavior*, *7*(2), 103-117.

McNab, B. K. (1997). On the utility of uniformity in the definition of basal rate of metabolism. *Physiological Zoology*, *70*(6), 718-720.

Mellor, D. J., Beausoleil, N. J., Littlewood, K. E., McLean, A. N., McGreevy, P. D., Jones, B., & Wilkins, C. (2020). The 2020 five domains model: Including human–animal interactions in assessments of animal welfare. *Animals*, *10*(10), 1870.

Mendl, M., Neville, V., & Paul, E. S. (2022). Bridging the gap: Human emotions and animal emotions. *Affective Science*, *3*(4), 703-712.

Mengoli, M., Oliva, J. L., Mendonça, T., Chabaud, C., Arroub, S., Lafont-Lecuelle, C., ... & Bienboire-Frosini, C. (2021). Neurohormonal profiles of assistance dogs compared to pet dogs: what is the impact of different lifestyles?. *Animals*, *11*(9), 2594.

Mitsui, S., Yamamoto, M., Nagasawa, M., Mogi, K., Kikusui, T., Ohtani, N., & Ohta, M. (2011). Urinary oxytocin as a noninvasive biomarker of positive emotion in dogs. *Hormones and behavior*, *60*(3), 239-243.

Mocharla, H., Mocharla, R., & Hodes, M.E. (1990) α-Amylase gene transcription in tissues of normal dog. *Nucleic acids research, 18*(4), 1031-1036.

Monaghan, P., Metcalfe, N. B., & Torres, R. (2009). Oxidative stress as a mediator of life history trade-offs: mechanisms, measurements and interpretation. *Ecology letters*, *12*(1), 75-92.

Mota-Rojas, D., Martínez-Burnes, J., Casas-Alvarado, A., Gómez-Prado, J., Hernández-Ávalos, I., Domínguez-Oliva, A., ... & Pereira, A. M. (2022). Clinical usefulness of infrared thermography to detect sick animals: Frequent and current cases. *CABI Reviews*, (2022).

Nagasawa, M., Kikusui, T., Onaka, T., & Ohta, M. (2009). Dog's gaze at its owner increases owner's urinary oxytocin during social interaction. *Hormones and behavior*, *55*(3), 434-441.

Nagasawa, M., Mitsui, S., En, S., Ohtani, N., Ohta, M., Sakuma, Y., ... & Kikusui, T. (2015). Oxytocin-gaze positive loop and the coevolution of human-dog bonds. *Science*, *348*(6232), 333-336.

Nater, U.M., La Marca, R., Florin, L., Moses, A., Langhans, W., Koller, M.M. & Ehlert, U. (2006). Stress-induced changes in human salivary alpha-amylase activity—associations with adrenergic activity. *Psychoneuroendocrinology, 31*(1), 49-58.

Neethirajan, S. (2024). Artificial intelligence and sensor innovations: enhancing livestock welfare with a human-centric approach. *Human-Centric Intelligent Systems*, *4*(1), 77-92.

Ogi, A., & Gazzano, A. (2023). Biomarkers of Stress in Companion Animals. *Animals*, *13*(4), 660.

Oken, B. S., Chamine, I., & Wakeland, W. (2015). A systems approach to stress, stressors and resilience in humans. *Behavioural brain research*, *282*, 144-154.

Oliva, J. L., Rault, J. L., Appleton, B., & Lill, A. (2015). Oxytocin enhances the appropriate use of human social cues by the domestic dog (Canis familiaris) in an object choice task. *Animal cognition*, *18*, 767-775.

Onaka, T. (2004). Neural pathways controlling central and peripheral oxytocin release during stress. *Journal of neuroendocrinology*, *16*(4), 308-312.

Palestrini, C., Previde, E. P., Spiezio, C., & Verga, M. (2005). Heart rate and behavioural responses of dogs in the Ainsworth's Strange Situation: A pilot study. *Applied Animal Behaviour Science*, *94*(1-2), 75-88.

Passantino, A., Quartarone, V., Pediliggeri, M. C., Rizzo, M., & Piccione, G. (2014). Possible application of oxidative stress parameters for the evaluation of animal welfare in sheltered dogs subjected to different environmental and health conditions. *Journal of Veterinary Behavior*, *9*(6), 290-294.

Paul, E. S., Browne, W., Mendl, M. T., Caplen, G., Held, S., Trevarthen, A., & Nicol, C. J. (2023). Affective trajectories: Are hens influenced by positive and negative changes in their living conditions?. *Applied Animal Behaviour Science*, *261*, 105883.

Payne, E., Bennett, P. C., & McGreevy, P. D. (2015). Current perspectives on attachment and bonding in the dog–human dyad. *Psychology research and behavior management*, 71-79.

Perry, E., Gulson, N., Liu Cross, T.-W., & Swanson, K.S. (2017). Physiological effects of stress related to helicopter travel in Federal Emergency Management Agency search-and-escue canines. *Journal of Nutritional Science, 6,* e28. Doi:10.1017/jns.2017.25

Polgár, Z., Blackwell, E. J., & Rooney, N. J. (2019). Assessing the welfare of kennelled dogs—A review of animal-based measures. *Applied animal behaviour science*, *213*, 1-13.

Posluns, J. A., Anderson, R. E., & Walsh, C. J. (2017). Comparing two canine personality assessments: Convergence of the MCPQ-R and DPQ and consensus between dog owners and dog walkers. *Applied Animal Behaviour Science*, *188*, 68-76.

Protopopova, A. (2016). Effects of sheltering on physiology, immune function, behavior, and the welfare of dogs. *Physiology & Behavior*, *159*, 95-103.

Richter, T., & von Zglinicki, T. (2007). A continuous correlation between oxidative stress and telomere shortening in fibroblasts. *Experimental gerontology*, *42*(11), 1039-1042.

Romero, T., Nagasawa, M., Mogi, K., Hasegawa, T., & Kikusui, T. (2014). Oxytocin promotes social bonding in dogs. *Proceedings of the National Academy of Sciences*, *111*(25), 9085-9090.

Romero, T., Nagasawa, M., Mogi, K., Hasegawa, T., & Kikusui, T. (2015). Intranasal administration of oxytocin promotes social play in domestic dogs. *Communicative & integrative biology*, *8*(3), e1017157.

Romero, L. M., & Beattie, U. K. (2022). Common myths of glucocorticoid function in ecology and conservation. *Journal of Experimental Zoology Part A: Ecological and Integrative Physiology*, *337*(1), 7-14.

Saarela, S., & Reiter, R. J. (1994). Function of melatonin in thermoregulatory processes. *Life sciences*, *54*(5), 295-311.

Schatz, S, & Palme, R. (2001). Measurement of faecal cortisol metabolites in cats and dogs: a non-invasive method for evaluating adrenocortical function. *Veterinary Research Communications, 25*(4), 271-87.

Schmidt-Nielsen, K., & Knut, S. N. (1984). *Scaling: why is animal size so important?*. Cambridge university press.

Segerstrom, S. C., & Miller, G. E. (2004). Psychological stress and the human immune system: a meta-analytic study of 30 years of inquiry. *Psychological bulletin*, *130*(4), 601.

Skandakumar, S., Stodulski, G., & Hau, J. (1995). Salivary IgA: A possible stress marker in dogs. *Animal Welfare*, *4*(4), 339-350.

Sohal, R. S., & Orr, W. C. (2012). The redox stress hypothesis of aging. *Free Radical Biology and Medicine*, *52*(3), 539-555.

Speakman, J. R., & Ward, S. (1998). Infrared thermography: principles and applications. *Zoology*, *101(3)*, 224-232.

Srithunyarat, T., Hagman, R., Höglund, O. V., Olsson, U., Stridsberg, M., Jitpean, S., ... & Pettersson, A. (2017). Catestatin and vasostatin concentrations in healthy dogs. *Acta Veterinaria Scandinavica*, *59*, 1-8.

Srithunyarat, T., Hagman, R., Höglund, O. V., Stridsberg, M., Hanson, J., Lagerstedt, A. S., & Pettersson, A. (2018). Catestatin, vasostatin, cortisol, and visual analog scale scoring for stress assessment in healthy dogs. *Research in veterinary science*, *117*, 74-80.

Starling, M. J., Branson, N., Thomson, P. C., & McGreevy, P. D. (2013). "Boldness" in the domestic dog differs among breeds and breed groups. *Behavioural processes*, *97*, 53-62.

Stewart, M., Webster, J. R., Schaefer, A. L., Cook, N. J., & Scott, S. L. (2005). Infrared thermography as a non-invasive tool to study animal welfare. *Animal Welfare*, *14*(4), 319-325.

Stewart, M., Webster, J. R., Verkerk, G. A., Schaefer, A. L., Colyn, J. J., & Stafford, K. J. (2007). Non-invasive measurement of stress in dairy cows using infrared thermography. *Physiology & behavior*, *92*(3), 520-525.

Stockham, S.L., Keeton, K.S. & Szladovits, B., (2003) Clinical assessment of leukocytosis: distinguishing leukocytoses caused by inflammatory, glucocorticoid, physiologic, and leukemic disorders or conditions. *Veterinary Clinics of North America Small Animal Practice*, *33*, 1335-1357.

Strasser, A., Niedermüller, H., Hofecker, G., & Laber, G. (1993). The effect of aging on laboratory values in dogs. *Journal of Veterinary Medicine Series A, 40*(1-10), 720-730.

Suh, B. Y., Liu, J. H., Rasmussen, D. D., Gibbs, D. M., Steinberg, J., & Yen, S. S. (1986). Role of oxytocin in the modulation of ACTH release in women. *Neuroendocrinology*, *44*(3), 309-313.

Sung, K. C., Lee, M. Y., Kim, Y. H., Huh, J. H., Kim, J. Y., Wild, S. H., & Byrne, C. D. (2018). Obesity and incidence of diabetes: Effect of absence of metabolic syndrome, insulin resistance, inflammation and fatty liver. *Atherosclerosis*, *275*, 50-57.

Svobodová, I., Chaloupková, H., Končel, R., Bartoš, L., Hradecká, L., & Jebavý, L. (2014). Cortisol and secretory immunoglobulin a response to stress in German shepherd dogs. *PLoS one*, *9*(3), e90820.

Travain, T., Colombo, E. S., Heinzl, E., Bellucci, D., Previde, E. P., & Valsecchi, P. (2015). Hot dogs: Thermography in the assessment of stress in dogs (Canis familiaris)—A pilot study. *Journal of veterinary behavior*, *10*(1), 17-23.

Travain, T., & Valsecchi, P. (2021). Infrared thermography in the study of animals' emotional responses: A critical review. *Animals*, *11*(9), 2510.

van der Laan, J.E., Vinke, C.M. & Arndt, S.S. (2022) Evaluation of hair cortisol as an indicator of long-term stress responses in dogs in an animal shelter and after subsequent adoption. *Scientific Reports 12*, 5117.

Venable, E.B., Bland,S.D., Holscher, H.D. & Swanson, K.S. (2016). Effects of air travel stress on the canine microbiome: A pilot study. *International Journal of Veterinary Health Science & Research 4(6),* 132-139.

Vera, F., Zenuto, R., & Antenucci, C. D. (2017). Expanding the actions of cortisol and corticosterone in wild vertebrates: a necessary step to overcome the emerging challenges. *General and comparative endocrinology*, *246*, 337-353.

Vít, M., Kučera, J., Lenárt, P., Novák, J., Zlámal, F., Reguli, Z., Bugala, M., Čihounková, J., Přecechtěl, P., Malčík, V., Vojtíšek, T., Kučerová, J.F., Eclerová, V., Tomandlová, M., Šíp, R., Ráčková, L., Grulichová, M., Tomandl, J. & Bienertová-Vašků, J. (2023). Biological factors and self-perception of stress in relation to freeze-like response in humans. *Psychoneuroendocrinology, 158*, 106382.

Von Borell, E., Langbein, J., Després, G., Hansen, S., Leterrier, C., Marchant-Forde, J., ... & Veissier, I. (2007). Heart rate variability as a measure of autonomic regulation of cardiac activity for assessing stress and welfare in farm animals—A review. *Physiology & behavior*, *92*(3), 293-316.

Webber, S., Cobb, M. L., & Coe, J. (2022). Welfare through competence: A framework for animal-centric technology design. *Frontiers in veterinary science*, *9*, 885973.

Webster, J. (2016). Animal welfare: Freedoms, dominions and "a life worth living". *Animals*, *6*(6), 35.

Zan, R. S., Rolinski, Z., Kowalski, C. J., Bojarska-Junak, A., & Madany, J. (2013). Diurnal and seasonal changes in endogenous melatonin levels in the blood plasma in dogs. *Polish Journal of Veterinary Sciences*, *16*(4).

Zapata, I., Serpell, J.A. & Alvarez, C.E. (2016). Genetic mapping of canine fear and aggression. *BMC Genomics 17*, 572.

Zhao, Z., Cao, J., Niu, C., Bao, M., Xu, J., Huo, D., ... & Speakman, J. R. (2022). Body temperature is a more important modulator of lifespan than metabolic rate in two small mammals. *Nature Metabolism*, *4*(3), 320-326.

Zupan, M., Buskas, J., Altimiras, J., & Keeling, L. J. (2016). Assessing positive emotional states in dogs using heart rate and heart rate variability. *Physiology & Behavior*, *155*, 102-111.